\documentclass[amsmath,amssymb,11pt]{article}
\usepackage{jheppub2}
\usepackage{xcolor}
\pdfoutput=1

\usepackage{amsmath,amssymb,amsthm,mathrsfs,bbm,bm}
\usepackage{latexsym,amscd,amsbsy,amsfonts,dsfont}
\usepackage{multirow}
\usepackage{graphicx}
\usepackage{xcolor}
\usepackage{caption}
\usepackage{comment}

\newcommand{\beq}{\begin{equation}}
\newcommand{\eeq}{\end{equation}}
\newcommand{\beqa}{\begin{eqnarray}}
\newcommand{\eeqa}{\end{eqnarray}}
\newcommand{\bea}{\begin{eqnarray}}
\newcommand{\eea}{\end{eqnarray}}

\newcommand{\ie}{{i.e.,\ }}

\newcommand{\sR}{\mathsf{R}}

\newcommand*{\affmark}[1][*]{\textsuperscript{#1}} 

\numberwithin{equation}{section}  
\allowdisplaybreaks

\title{The Relaxation of Quantum Near-Extremal Black Holes

\vspace{10pt}
\Large{(I) Microcanonical States, Pruning of Quasinormal Modes, and the Aretakis Instability}
}

\author{Roberto Emparan\affmark[1,2],}
\emailAdd{emparan@ub.edu}
\author{Sameer Murthy\affmark[3],}
\emailAdd{sameer.murthy@kcl.ac.uk}
\author{Stefano Trezzi\affmark[2]}
\emailAdd{strezzi@icc.ub.edu}

\affiliation{
\affmark[1]Institució Catalana de Recerca i Estudis Avançats (ICREA),
 Passeig Lluis Companys, 23, 08010 Barcelona, Spain\\
\affmark[2]Departament de Física Quàntica i Astrofísica and
  Institut de Ciències del Cosmos (ICCUB), Martì i Franquès~1,
 Universitat de Barcelona, 08028 Barcelona, Spain\\
\affmark[3]Department of Mathematics, King's College London, The Strand, London WC2R 2LS, UK\\
}

\abstract{
The late-time relaxation of a near-extremal black hole is classically governed by an infinite tower of purely damped quasinormal modes. This poses a puzzle: the higher overtones decay at rates much larger than the energy $E$ of the black hole above extremality, so they would need black hole states below the extremal mass. We resolve it by including the quantum fluctuations that become large when $E$ falls below a scale $E_b$. We compute the retarded Green's function of a probe of conformal dimension $\Delta$ in a black hole of fixed energy, and find that each classical mode splits into a pair of poles with real parts set by $E_b$. The perturbed black hole decays as an unstable quantum system, and a pair is a resonance, i.e., a quasinormal mode, only if its energy lies within the accessible black hole spectrum. This prunes the tower to finitely many modes, all damped faster than $\Delta^2E_b$, and below $E=\Delta^2E_b/2$ none is left.
For generic $\Delta$, the response of the black hole decays first as $t^{-2\Delta}$ and finally as $t^{-3/2}$, with an exponential stage in between if $E\gg\Delta^2E_b$.

The same quantum effects bear on the Aretakis instability of extremal black holes, in which transverse derivatives of a perturbation grow without bound on the horizon. Like the tower of modes, it originates in the scale invariance of the classical throat, which the quantum fluctuations break.
The $t^{-2\Delta}$ decay, which is the boundary counterpart of the Aretakis growth, ends by $t\sim1/E_b$, when the throat ceases to be classical, and by then horizon derivatives have grown only to a finite value. Their unbounded growth is an artifact of taking the semiclassical limit before the late-time one, two limits that in quantum mechanics do not commute.
}

\begin{document}


\maketitle

\section{Introduction and Summary}
\label{sec:intro}

How does a quantum black hole relax?

For a classical black hole the answer has long been known: at late times a perturbation decays as a superposition of exponentially damped sinusoids, the quasinormal modes, whose 
frequencies depend only on the mass, charge and spin of the black hole \cite{Vishveshwara:1970zz,Press:1971wr,Berti:2009kk}.\footnote{We set the classical spin of the black hole to zero, so near-extremality comes from the charge only.}
In asymptotically flat space the ringing is eventually overtaken by a power-law tail, from backscattering off the Newtonian potential far from the hole \cite{Price:1971fb,Arnaudo:2025kit}.

Near extremality this picture acquires a new and delicate element. A long throat, with geometry close to AdS$_2\times S^2$, stretches out from the horizon, and it carries its own characteristic quasinormal modes. For a perturbation that propagates in the throat as a field of conformal dimension $\Delta$, these modes lie at
\begin{equation}\label{eq:zdm}
    \omega_n=-2\pi i T(\Delta+n)\,, \qquad n=0,1,2,\dots.
\end{equation}
They are purely damped, with rates set by the temperature \cite{Zimmerman:2015trm,Kapec:2024zdj,LoChiatto:2026tne}. The tower is infinite, and its rates grow without limit. As $T\to 0$ the tower becomes long-lived and dense, and at extremality it condenses into a branch point at zero frequency. The exponential decay then turns into a power law, $t^{-2\Delta}$, the imprint of a throat that has become infinitely long and scale invariant \cite{Gralla:2018xzo,Hadar:2018izi,Arnaudo:2024sen}.

In the extremal limit, however, classical relaxation fails: the black hole never fully settles down after it is perturbed. Outside the horizon the perturbation decays together with all its derivatives. On the horizon the perturbation itself still decays, but its $N$-th transverse derivative behaves as $v^{N-\Delta}$, where $v$ is the advanced time, an affine parameter along the null generators \cite{Gralla:2018xzo}. Those with $N > \Delta$ therefore grow without bound \cite{Aretakis:2011ha,Aretakis:2011hc,Aretakis:2012ei,Lucietti:2012xr,Lucietti:2012sf,Murata:2013daa}. This is the Aretakis instability. Its growth is confined to the horizon, but its origin lies in the throat:
\begin{itemize}
\item it follows from the scaling symmetry of AdS$_2$ \cite{Gralla:2018xzo};
\item it is the horizon counterpart of the $t^{-2\Delta}$ decay at the mouth of the throat \cite{Gralla:2018xzo,Hadar:2018izi};
\item nonlinear evolutions find Aretakis-like growth even when no horizon forms \cite{Gelles:2026bix};
\item slightly away from extremality, with a long but finite throat, it survives as a transient that grows until times of order $1/T$~\cite{Gralla:2016sxp,Zimmerman:2016qtn}.
\end{itemize}
The extremal limit thus produces a throat so long that perturbations never quite die in it.\footnote{The extremal limit here is taken along a family of black hole states, not as a physical process. Classically, collapse can form an exactly extremal black hole in finite time \cite{Kehle:2022uvc}, but in quantum gravity this state is never reached \cite{Turiaci:2026gtl}.}

It is now known, however, that this regime is never realized. Near extremality, the mass of the black hole exceeds its extremal value by an energy $E$ that grows quadratically with the temperature,
\begin{equation}\label{eq:ET}
    E=\frac{2\pi^2T^2}{E_b}\,,\qquad E_b=\frac{\pi}{r_0S_0}\,,
\end{equation}
where $r_0$ and $S_0$ are the horizon radius and the Bekenstein--Hawking entropy of the extremal Reissner--Nordstr\"om black hole.\footnote{For Reissner--Nordstr\"om--AdS$_4$ with AdS radius $L$, $E_b=\frac{\pi}{r_0S_0}\left(1+6r_0^2/L^2\right)$.}
The specific heat $C=dE/dT$ then decreases as the black hole cools, and when $E\lesssim E_b$ the energy fluctuations, $\delta E/E\sim 1/\sqrt{C}$, become of order one, so the classical geometry of the throat breaks down~\cite{Preskill:1991tb,Iliesiu:2020qvm}.

Relaxation hides a related puzzle, which to our knowledge has not been raised before. In quantum mechanics a mode that decays at a rate $\Gamma$ involves energies spread over a range of order $\Gamma$. The overtones of \eqref{eq:zdm} with $n\gg T/E_b$ decay at rates much larger than $E$, and when $E\ll E_b$ all the modes do. The classical decay requires states that the black hole does not have.

Quantum gravity is then unavoidable, but, remarkably, it is under analytic control. The large fluctuations come from a few soft modes of the throat, mainly the Schwarzian mode, whose quantum dynamics can be solved exactly~\cite{Maldacena:2016upp,Stanford:2017thb,Mertens:2017mtv,Mertens:2022irh}. Their effects on the thermodynamics, emission and absorption, tidal deformability, decoherence and transport in near-extremal black holes have been studied in detail in recent years
\cite{Iliesiu:2020qvm,Iliesiu:2022onk,Bai:2023hpd,Kapec:2023ruw,Rakic:2023vhv,Kapec:2024zdj,Brown:2024ajk,Maulik:2025hax,Emparan:2025sao,Biggs:2025nzs,Li:2025vcm,Emparan:2025qqf,Betzios:2025sct,Nian:2025oei,PandoZayas:2025snm,Bhattacharjee:2025wfv,Cremonini:2025yqe,Rakic:2025svg,Gouteraux:2025exs,Kanargias:2025vul,Luo:2026epp,Biggs:2026zlp,Maulik:2026dro,Das:2026ujn,Bhattacharjee:2026nmo,Cano:2026scd}

So quantum effects become large precisely where relaxation shows its most peculiar features. This gives us a rare opportunity to follow in detail the dynamics of a quantum geometry where it is richest and, in particular, to ask which of these modes the black hole can sustain, what replaces the condensing tower, and what becomes of the Aretakis instability.

\paragraph{Ensemble-dependent relaxation.}
How a quantum black hole relaxes depends on which state it is in. Classically this hardly matters: by ensemble equivalence, a black hole of given energy relaxes like a thermal one at the temperature \eqref{eq:ET}. In the quantum regime, instead, the large energy fluctuations mean that a thermal ensemble contains states with energies spread over a range as wide as their mean \cite{Biggs:2025nzs,Biggs:2026zlp}, so it is a different system from a state of fixed energy, and it relaxes differently. Both are natural: an isolated black hole has fixed energy, while a black hole in equilibrium with its surroundings, as in the thermal states of a holographic dual, has fixed temperature. In this paper we focus on states of fixed energy, and discuss what the Aretakis instability looks like for them. In a companion paper we turn to black holes at fixed temperature, where the broad spread of energies blurs the quasinormal modes and speeds up the late-time decay \cite{Emparan:2026relaxII}.

\paragraph{Fields in a quantum throat.}
Our probes are fields in the black hole background whose partial waves become, in the throat, fields in AdS$_2$ of conformal dimension $\Delta$. Each acts on the black hole through an operator $O_\Delta$ at the mouth. This covers a large class of fields, in asymptotically flat as well as AdS black holes, including the gravitational and electromagnetic perturbations of the black hole itself.
The simplest example is a massless neutral scalar: its partial wave of angular momentum $\ell$ in Reissner–Nordström–AdS$_4$ with AdS radius $L$ has a dimension given by
\begin{equation}\label{eq:Delta}
    \Delta(\Delta-1)=\frac{\ell(\ell+1)}{1+6r_0^2/L^2}\,,
\end{equation}
so $\Delta$ is not an integer for $\ell\geq1$ and generic $r_0/L$. We take $\Delta$ to be real and positive. Our exact results hold for any such $\Delta$, but when we make parametric estimates we take it to be of order one.\footnote{Integer values, as for a massless scalar in asymptotically flat Reissner--Nordstr\"om, are special in the classical throat, whose response then has no poles; quantum effects restore them (section~\ref{sec:time_dom_qu_response}). The response at real frequencies for integer $\Delta$ was obtained in \cite{Cano:2026scd}.}

The black hole is initially spinless, in a state of fixed energy $E$ above extremality. A partial wave of angular momentum $\ell$ drives transitions into the continuum of black hole states of spin $\ell$, which starts, for Reissner--Nordstr\"om, at the rotational energy \cite{Iliesiu:2020qvm}\footnote{In general $E_{0,\ell}=\ell(\ell+1)/(2I)$, where the moment of inertia $I$ gives the increase of the extremal mass with spin at fixed charge \cite{Iliesiu:2020qvm}. In Planck units, $I=r_0^3$ for Reissner--Nordstr\"om, while $I=r_0^3(L^2+2r_0^2)/(L^2+r_0^2)$ for Reissner--Nordstr\"om--AdS$_4$.}
\begin{equation}\label{eq:E0l}
    E_{0,\ell}=\frac{\ell(\ell+1)}{2}\,E_b\,.
\end{equation}
We compute the exact retarded Green's function of the field in this state, including the Schwarzian and rotational fluctuations of the throat \cite{Mertens:2017mtv,Brown:2024ajk,Cano:2026scd}. The transitions, by absorption or emission, occur at real frequencies where the final energy exceeds $E_{0,\ell}$. The final states form a continuum, and their contributions merge into branch cuts that start at its bottom, at the frequencies $\omega=\pm(E_{0,\ell}-E)$.
As for resonances in scattering theory, the poles of the Green's function lie on sheets reached by analytic continuation across the cuts.

\paragraph{Quasinormal spectrum: a pruned quantum Christmas tree.}
Performing this continuation we find that quantum effects split each classical mode \eqref{eq:zdm} into a pair of poles,
\begin{equation}\label{eq:qqnm}
    \omega_n^\pm=\pm\left(E_{0,\ell}-\frac{E_b}{2}(\Delta+n)^2\right)-i\sqrt{2E_bE}\,(\Delta+n)\,,\qquad n=0,1,2,\dots
\end{equation}
The damping is exactly that of the modes \eqref{eq:zdm} of a classical black hole with the same energy, but the frequencies now have real parts set by $E_b$, so they oscillate at a rate independent of the energy. In the complex frequency plane the pairs form a ``Christmas tree'' pattern, reminiscent of the quasinormal spectrum of black branes \cite{Berti:2009kk,Grozdanov:2018gfx}.

\begin{figure}[t]
    \centering
    \includegraphics[width=\textwidth]{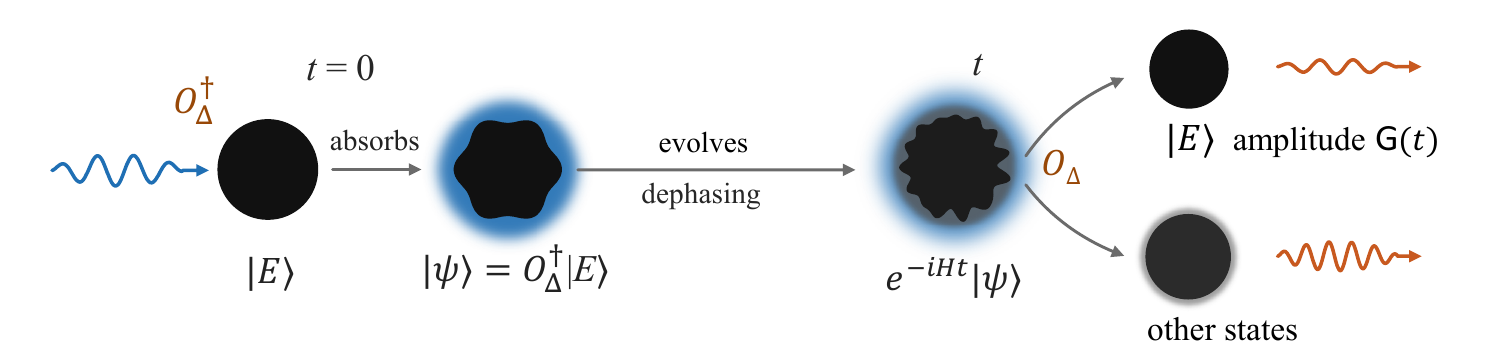}
    \caption{\small Measuring the relaxation of a quantum black hole. At time $0$, the black hole in the state $|E\rangle$ absorbs a quantum and is excited into $|\psi\rangle=O_\Delta^\dagger|E\rangle$, a superposition of black hole states with energies $E'$, which dephase as they evolve. At time $t$, the black hole re-emits the quantum and returns to $|E\rangle$ with amplitude $\mathsf{G}(t)$, the survival amplitude of $|\psi\rangle$; otherwise it ends in another state. Only the return to $|E\rangle$ contributes to the linear response. Stimulated emission gives the complex conjugate of $\mathsf{G}(t)$, and $G_R(t)=-2\,\mathrm{Im}\,\mathsf{G}(t)$. The dephasing makes $\mathsf{G}(t)$ decay: exponentially at the rates of the quasinormal modes, which are resonances in the continuum of black hole states, and as a power law at late times, set by the bottom of the continuum.}
    \label{fig:relax}
\end{figure}

The perturbed black hole then relaxes like an unstable quantum system \cite{Fonda:1978dk}. It is not in an energy eigenstate, but in the state $|\psi\rangle=O_\Delta^\dagger|E\rangle$, a superposition of black hole states of spin $\ell$, with energies $E'$, each with probability proportional to the weight
\begin{equation}\label{eq:weight}
    \mathsf{W}(E')=\rho(E')\,|\langle E'|O_\Delta|E\rangle|^2\,,
\end{equation}
where $\rho$ is their density. In the response the probe acts twice: at time $0$ it excites the black hole into $|\psi\rangle$, and at time $t$ it undoes the excitation, as the black hole re-emits the quantum (figure~\ref{fig:relax}). The black hole returns to $|E\rangle$ only to the extent that it is still in $|\psi\rangle$, so the amplitude for this is the overlap of $|\psi\rangle$ with itself after a time $t$, which is given by
\begin{equation}\label{eq:survival}
    \mathsf{G}(t)=\int_{E_{0,\ell}}^{\infty}dE'\,\mathsf{W}(E')\,e^{-i(E'-E)t}\,,
\end{equation}
also known as the survival amplitude of $|\psi\rangle$.\footnote{$\mathsf{G}(t)$ is the Wightman function, usually denoted as $G^>(t)$, in the state $|E\rangle$, and its Fourier transform at frequency $\omega$ is $2\pi\,\mathsf{W}(E+\omega)$. For $\Delta\geq1/2$ the integral converges only with $t\to t-i\epsilon$, $\epsilon\to0^+$.}
The retarded Green's function is $i$ times the difference between absorption, which takes the black hole to $E+\omega$ and gives $\mathsf{G}(t)$, and stimulated emission, which takes it to $E-\omega$ and gives its complex conjugate:
\begin{equation}
    G_R(t)=i\left[\mathsf{G}(t)-\mathsf{G}(t)^*\right]=-2\,\mathrm{Im}\,\mathsf{G}(t)\,,\qquad t>0\,.
\end{equation}
The energies of the black hole states are real, so each component of $\mathsf{G}(t)$ oscillates forever with constant amplitude, and none decays by itself. The decay is a collective phenomenon: the contributions of neighbouring energies drift out of phase and, for a continuous spectrum, eventually cancel. Complex frequencies appear when the same integral is rewritten by moving the contour below the real axis. The dephasing then becomes a sum of exponentials, each from a pole whose real and imaginary parts give its energy and decay rate, plus a contribution from the branch cuts, which decays as a power law. The poles $\omega_n^+$ come from $\mathsf{G}(t)$, and their mirror images $\omega_n^-=-(\omega_n^+)^*$ from its complex conjugate, so each pair describes a single resonance of the excited black hole. The resonances are the quasinormal modes. They are broad, and $\mathsf{W}(E')$ rises smoothly through their energies, with no peaks.\footnote{This is for $\Delta\geq 1/2$. For $\Delta<1/2$ the weight has a maximum, since it must fall at large $E'$, and as $\Delta\to0$ it narrows into a peak at the least damped mode; see section~\ref{sec:quantum_setup}.}

\begin{figure}[t]
    \centering
    \includegraphics[width=\textwidth]{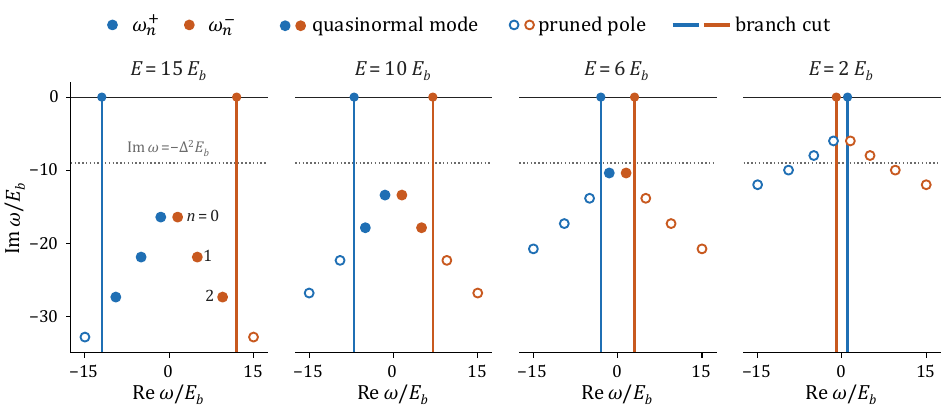}
    \caption{\small Pruning of the quasinormal spectrum as the energy above extremality $E$ is lowered, for a massless scalar with $\ell=2$ in Reissner--Nordstr\"om, so $\Delta=3$, $E_{0,\ell}=3E_b$. We draw the branch cuts running down from the branch points at $\omega=\pm(E_{0,\ell}-E)$, so that the plane shows the continuation of the response from the upper half-plane. As $E$ falls, the poles rise vertically, as $\sqrt{E}$, and the cuts move across them. Blue poles to the right of the blue cut and orange poles to the left of the orange cut (filled) lie on this continuation and are quasinormal modes. The others (hollow) lie behind their cut, on another sheet, and would take the black hole to energies below the continuum of spin-$\ell$ states. No quasinormal mode enters the region above the dotted line, $\mathrm{Im}\,\omega=-\Delta^2E_b$, and below $E=\Delta^2E_b/2$ (last panel) none is left.}
    \label{fig:pruning}
\end{figure}

But a resonance needs states to superpose. The $n$-th pair, by absorption or by stimulated emission, would take the black hole from the energy $E$ to $E+E_{0,\ell}-(\Delta+n)^2E_b/2$. The continuum starts at $E_{0,\ell}$, so the weight $\mathsf{W}$ vanishes at that energy unless
\begin{equation}\label{eq:prune}
    E>\frac{(\Delta+n)^2}{2}\,E_b\,,
\end{equation}
independently of the rotational energy. Otherwise the pole lies below the bottom of the continuum and leaves no separate imprint on the response. In terms of the decay rate, the condition \eqref{eq:prune} is $|\mathrm{Im}\,\omega_n^\pm|<2E$: a mode survives only if the black hole has the energy for its decay rate. Only the pairs at the top of the tree are therefore quasinormal modes, and a black hole of fixed energy has finitely many of them.\footnote{The poles $\omega_n^+$ for $\Delta=\ell+1$ have also been found in \cite{Bhattacharjee:2026nmo}, without determining which of them are resonances.} This resolves the puzzle of the infinite tower that we raised above. As the energy is lowered towards extremality, the poles move towards the real axis, but the tree is pruned from the bottom up. One by one, the pairs drop below the bottom of the continuum and cease to be quasinormal modes, each when its damping has come down to $(\Delta+n)^2E_b$. So no quasinormal mode ever comes close to the real axis, and below $E=\Delta^2E_b/2$ none is left (see figure~\ref{fig:pruning}).

Conversely, when $E\gg E_b$ the superposition extends only a few $T$ below the initial energy, and the bottom of the continuum lies far below. The condition \eqref{eq:prune} then removes only very high overtones, with $n\gtrsim\sqrt{2E/E_b}$, which matter only at times $t\lesssim1/E$ where the decay is conformal.
We then recover the classical black hole, with its tower of modes \eqref{eq:zdm}, until $t\sim1/E_b$, when quantum effects make themselves felt.

This picture follows from the sheet structure of the Green’s function in the complex frequency plane, which we study in section~\ref{sec:quantum_Green}.

\begin{figure}[t]
    \centering
    \includegraphics[width=.95\textwidth]{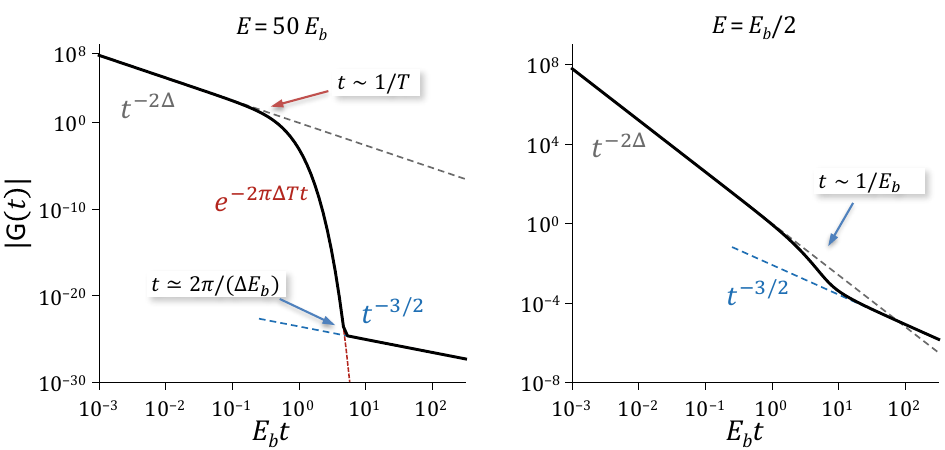}
    \caption{\small Decay in time of the survival amplitude $|\mathsf{G}(t)|$, in units of $(2\ell+1)E_b^{2\Delta}$, for $\Delta=1.3$ and $E_{0,\ell}=0$, computed from the exact result of section~\ref{sec:time_dom_qu_response}. The response $G_R=-2\,\mathrm{Im}\,\mathsf{G}$ has the same envelope. Left: when $E\gg E_b$, the conformal decay $t^{-2\Delta}$ (grey) gives way at $t\sim1/T$ to the exponential decay of the quasinormal modes, as in the classical throat (red), until the power law $t^{-3/2}$ (blue) emerges at $t\simeq2\pi/(\Delta E_b)$.
    Right: below $E=\Delta^2E_b/2$ there are no quasinormal modes, and the conformal decay crosses over at $t\sim1/E_b$ to the $t^{-3/2}$ tail, with no intermediate exponential stage. After that time, the throat is no longer classical.}
    \label{fig:stages}
\end{figure}

\paragraph{Relaxation in stages.}
In the time domain (section~\ref{sec:time_dom_qu_response}), the response to a perturbation decays in stages that depend on the energy of the black hole:
\begin{itemize}
\item When $E\gg E_b$, the state has a well-defined temperature $T$, related to its energy by \eqref{eq:ET}, and there are three stages (figure~\ref{fig:stages}):
\begin{itemize}
\item at early times, $t\ll 1/T$, the conformal law $t^{-2\Delta}$ of the throat;\footnote{For integer $\Delta$ the early decay of the response is generically $t^{1-2\Delta}$; see section~\ref{sec:time_dom_qu_response}.}
\item at intermediate times, $1/T\ll t\ll 1/E_b$, the decay $e^{-2\pi\Delta Tt}$ of the least damped mode;
\item at late times, $t\gg 1/E_b$, a power law $t^{-3/2}$.
\end{itemize}
\item Below $E=\Delta^2E_b/2$ (the fully quantum regime $E\lesssim E_b$) there is no exponential decay stage:
\begin{itemize}
\item at early times, $t\ll 1/E_b$, the conformal law $t^{-2\Delta}$;
\item at late times, $t\gg 1/E_b$, the power law $t^{-3/2}$.
\end{itemize}
\end{itemize}
At energies in between, there are quasinormal modes and they show in the response, but as damped exponentials with hardly any oscillation.

The late-time power law comes from the bottom edge of the continuum of black hole states. Its exponent is fixed by the square-root density of states at this edge, so it does not depend on $\Delta$, and it oscillates at the frequency $|E-E_{0,\ell}|$ of the transition to the edge. An unstable quantum state also departs from exponential decay at late times, for the same reason: its spectrum is bounded from below \cite{Khalfin:1958,Winter:1961zz,Fonda:1978dk,Rothe:2006rma}. The power law is present also for black holes in the semiclassical regime: at late enough times any black hole becomes sensitive to the quantum fluctuations of its throat.

All of this holds in the disk approximation of the gravitational path integral, which includes the quantum fluctuations of the throat exactly and misses effects at energies and inverse times exponentially small in the entropy. In asymptotically flat space, Hawking emission ends the tail earlier, but only after times longer than $1/E_b$ by powers of the entropy (section~\ref{sec:concl}).

\paragraph{The fate of the Aretakis instability.}

Classically, the $t^{-2\Delta}$ decay is the boundary image of a self-similar solution with an unbounded growth of the horizon derivatives, which is the Aretakis instability.
As we show in sections~\ref{sec:quantum_Green} and~\ref{sec:time_dom_qu_response}, quantum effects remove the zero-damped quasinormal modes, and at late times they replace the $t^{-2\Delta}$ decay with the $t^{-3/2}$ tail. This suggests that the Aretakis instability disappears as well.

However, the absence of these modes in the quantum theory does not fully settle the question, since our analysis is based on boundary correlators, while the Aretakis growth occurs on the horizon.
Scaling symmetry connects the boundary and the horizon in the classical theory, but this symmetry is broken by quantum effects. The scale that breaks it is~$E_b$, and however close the black hole is to extremality, its throat is scale invariant only over a finite range of times.

The $t^{-2\Delta}$ decay thus lasts only until $t\sim 1/E_b$, and the self-similar solution cannot hold beyond that time. It would persist at all times only if $E_b\to0$ were taken before $t\to\infty$, but this order is physically wrong: $E_b$ is fixed by the charge of the black hole, and close enough to extremality every black hole has $E\lesssim E_b$.
It is then natural to expect that the Aretakis growth comes to an end: the horizon derivatives can grow only to a finite value before the throat ceases to be classical.

Still, at late times the response decays as a different power law, $t^{-3/2}$. Whether this drives some other kind of growth in the fully quantum regime is a question we leave open, since it must first be rephrased, for instance in terms of what an observer falling into the quantum throat at late times measures \cite{Franken:2026bff,Franken:2026lgp}.

\paragraph{Plan.}
Section~\ref{sec:classical_Green} describes the relaxation of the classical throat in terms that carry over to the quantum one. In section~\ref{sec:quantum_setup} we set up the quantum problem and describe the perturbed black hole as an unstable state. Section~\ref{sec:quantum_Green} continues the response function to complex frequencies, determines which of the poles are quasinormal modes, and examines the classical and low-energy limits. Section~\ref{sec:time_dom_qu_response} gives the response in time, as the quasinormal modes plus an integral around the cuts, and derives from it the stages of the relaxation. Section~\ref{sec:particle} gives a picture of the relaxation as the motion of a particle whose position is the length of the geodesic through the throat, and section~\ref{sec:quantum_Aretakis} discusses what becomes of the Aretakis instability.  We conclude in section~\ref{sec:concl}. Appendix~\ref{app:Stieltjes} gives a closed form of the Stieltjes transform of the weight, and appendix~\ref{app:particle} the exact formulation of the particle picture.

\section{Classical relaxation in the throat}
\label{sec:classical_Green}

We begin by examining how the classical throat of a near-extremal black hole relaxes when perturbed. We approach the problem in a way that will carry over to the quantum case: through the response function at the mouth of the throat, its singularities in the complex frequency plane, and the decay they produce in time.

\subsection{The geometry and fields in it}
\label{subsec:throat}

The Reissner--Nordstr\"om--AdS$_4$ black hole with outer and inner horizons at $r_+$ and $r_-$ is
\begin{equation}\label{eq:RNAdS}
    ds^2=-f(r)\,dt^2+\frac{dr^2}{f(r)}+r^2d\Omega^2\,,
\end{equation}
with
\begin{equation}\label{eq:fRN}
    f(r)=\frac{(r-r_+)(r-r_-)}{r^2}\left[1+\frac{r^2+(r_++r_-)\,r+r_+^2+r_+r_-+r_-^2}{L^2}\right],
\end{equation}
and the asymptotically flat Reissner--Nordstr\"om solution is recovered when $L\to\infty$. We use $r_\pm$ as the parameters of the solution, instead of the mass $M$ and the charge $Q$.\footnote{With $f(r)=1-\frac{2M}{r}+\frac{Q^2}{r^2}+\frac{r^2}{L^2}$, they are
\begin{equation*}
    M=\frac{r_++r_-}{2}\left(1+\frac{r_+^2+r_-^2}{L^2}\right),\qquad Q^2=r_+r_-\left(1+\frac{r_+^2+r_+r_-+r_-^2}{L^2}\right).
\end{equation*}}
At extremality the two horizons coincide at $r_0$. Close to it, when $r_+-r_-\ll r_0$, and in the region near the horizon where $r-r_+\ll r_0$,
the factor in square brackets is constant and\footnote{Throughout, $\simeq$ gives the leading behaviour of the left-hand side when a parameter becomes asymptotically small or large, so that $f(x) \simeq g(x)$ as $x \to x_0$ means $\lim_{x \to x_0} f(x) / g(x) = 1$.}
\begin{equation}\label{eq:fthroat}
    f(r)\simeq\frac{(r-r_+)(r-r_-)}{L_2^2}\,,\qquad L_2=\frac{r_0}{\sqrt{1+6r_0^2/L^2}}\,.
\end{equation}
In terms of the temperature of the black hole and a new radial coordinate,
\begin{equation}\label{eq:TR}
    T=\frac{r_+-r_-}{4\pi L_2^2}\,,\qquad R=\frac{r-r_+}{L_2^2}\,,
\end{equation}
the geometry in this region becomes
\begin{equation}\label{eq:throat}
    ds^2=L_2^2\left[-R(R+4\pi T)\,dt^2+\frac{dR^2}{R(R+4\pi T)}\right]+r_0^2\,d\Omega^2\,.
\end{equation}
This is the throat: AdS$_2\times S^2$, with AdS$_2$ of radius $L_2$ in the coordinates of a black hole of temperature $T$ and event horizon at $R=0$. The time $t$ is still that of the asymptotic observers, so frequencies and decay rates computed in the throat are the ones they measure.

The throat connects to the rest of the geometry through its mouth,
\begin{equation}\label{eq:mouth}
    4\pi T\ll R\ll \frac{r_0}{L_2^2}\,.
\end{equation}
For perturbations of frequency $|\omega|\ll r_0/L_2^2$, the mouth plays the role of the boundary of AdS$_2$, at $R\to\infty$, where the perturbations are sourced and their response is read.
We take $\omega$ and $T$ small compared with $r_0/L_2^2$, keeping the ratio $\omega/T$ arbitrary.

A partial wave of a neutral field, with angular momentum numbers $\ell$ and $m$, becomes in the throat a field $\Psi_{\ell m}(t,R)$ in AdS$_2$. It acquires an effective mass $\mu$ from the angular momentum on the sphere, which fixes its conformal dimension $\Delta$,\footnote{We consider only fields satisfying the Breitenlohner--Freedman bound, $\mu^2L_2^2\geq-1/4$, so that the conformal dimension $\Delta$ is real.}
\begin{equation}\label{eq:mDelta}
    \mu^2L_2^2=\Delta(\Delta-1)\,,
\end{equation}
where we take the larger of the two roots for $\Delta$. For a massless scalar field in four dimensions, $\mu^2=\ell(\ell+1)/r_0^2$, and this reproduces \eqref{eq:Delta}. The field need not be
external
to the theory: the perturbations of the metric and gauge field of the black hole are also of this kind. In Reissner--Nordstr\"om they are coupled, but in the throat they combine into fields with integer $\Delta$ \cite{Zimmerman:2015trm}, which also show the Aretakis instability \cite{Lucietti:2012xr}.

At the mouth, where $R\to\infty$, a field of dimension $\Delta$ behaves as
\begin{equation}\label{eq:AB}
    \Psi_{\ell m}\to A_{\ell m}(t)\,R^{\Delta-1}+B_{\ell m}(t)\,R^{-\Delta}\,.
\end{equation}
The first term dominates towards the mouth, and is fixed there by the perturbation that is applied from outside.
The second term decays, and is the response that the throat develops.
At the low frequencies we consider they are not ingoing and outgoing waves: in the exterior they match onto the applied field and the multipole that it induces, as in a tidal deformation. We therefore take $A_{\ell m}$ as the source and $B_{\ell m}$ as the response, as is usual at the boundary of an asymptotically AdS space.\footnote{This is the standard quantization. When $-1/4<\mu^2L_2^2<0$ the term $R^{\Delta-1}$ also decays, and the alternative quantization, with $B_{\ell m}$ as the source and operator dimension $1-\Delta$, is also possible. The matching to the exterior geometry decides between the two~\cite{Emparan:2025qqf,Kanargias:2025vul,Cano:2026scd}. Our results then hold with $\Delta\to1-\Delta$.}

It is convenient to describe this in the language of AdS/CFT, where the throat is dual to a quantum mechanical system at its boundary, the mouth, and the classical geometry \eqref{eq:throat} describes this system in a thermal state. Each partial wave of the field is dual to an operator $O_{\Delta\ell m}$ of dimension $\Delta$, and the source acts on the system by adding to its action the coupling
\begin{equation}\label{eq:coupling}
    I_{\rm source}=\sum_{m=-\ell}^{\ell}\int_{-\infty}^{\infty}dt\,A_{\ell m}(t)\,O^\dagger_{\Delta\ell m}(t)\,.
\end{equation}
The response to the source is then governed by the retarded Green's function of these operators,
\begin{equation}\label{eq:GRdef}
    G_R^{\rm cl}(t)=i\,\Theta(t)\sum_{m=-\ell}^{\ell}\left\langle\left[O_{\Delta\ell m}(t),O^\dagger_{\Delta\ell m}(0)\right]\right\rangle\,,
\end{equation}
where~$\Theta$ is the Heaviside-theta function, and
the expectation value is taken in the state of the throat, here the classical black hole of temperature $T$. We sum over $m$, which by spherical symmetry gives a factor $2\ell+1$. The Fourier transform of $G_R^{\rm cl}$,
\begin{equation}\label{eq:GRFT}
    \widehat G_R^{\rm cl}(\omega)=\int_{-\infty}^{\infty}dt\,e^{i\omega t}\,G_R^{\rm cl}(t)\,,
\end{equation}
is analytic for $\mathrm{Im}\,\omega>0$.
The normalization of the operators is fixed at short times, $t\ll1/T$, when the temperature does not matter and the throat is scale invariant, so their two-point function is a power law. We choose
\begin{equation}\label{eq:Onorm}
    \left\langle O_{\Delta\ell m}(t)\,O^\dagger_{\Delta\ell m}(0)\right\rangle \simeq (it)^{-2\Delta}=e^{-i\pi\Delta}\,t^{-2\Delta}\,,\qquad 0 < t \ll \frac{1}{T}\,,
\end{equation}
which is the continuation to real time, $t=-i\tau$, of the correlator $\tau^{-2\Delta}$ in Euclidean time $\tau>0$.

\subsection{Response function and quasinormal modes}
\label{subsec:classical_QNMs}

For a field $\Psi=e^{-i\omega t}\psi(R)$, where we drop the angular indices, the wave equation in the metric \eqref{eq:throat} is
\begin{equation}\label{eq:radial}
    \partial_R\left[R(R+4\pi T)\,\partial_R\psi\right]+\left[\frac{\omega^2}{R(R+4\pi T)}-\Delta(\Delta-1)\right]\psi=0\,,
\end{equation}
whose solutions at large $R$ are the two powers in \eqref{eq:AB}.
In terms of
\begin{equation}\label{eq:xdef}
    x=\frac{R}{2\pi T} \,,
\end{equation}
\eqref{eq:radial} is the Legendre equation,
whose solutions are linear combinations of $P^{\,i\omega/2\pi T}_{\Delta-1}(1+x)$ and $Q^{\,i\omega/2\pi T}_{\Delta-1}(1+x)$.
Three ingredients determine the shape of the response.

The first is the condition at the horizon. A retarded response must be ingoing at the future horizon \cite{Son:2002sd}, and this selects the solution $P$. Near $x=0$ it behaves as
\begin{equation}\label{eq:ingoing}
    \psi\propto x^{-i\omega/4\pi T}=e^{-i\omega r_*}\,,\qquad r_*=\frac{1}{4\pi T}\ln x\,,
\end{equation}
where $r_*$ is the tortoise coordinate near the horizon, so that $\Psi\propto e^{-i\omega(t+r_*)}$ depends only on the advanced time.

The second is the behaviour of this solution at the mouth. Expanding $P$ at large $x$ and comparing with \eqref{eq:AB}, the Fourier transforms of the source and the response are
\begin{equation}\label{eq:ABcl}
    \widehat A(\omega)=\frac{(\pi T)^{1-\Delta}\,\Gamma\!\left(\Delta-\frac12\right)}{\Gamma\!\left(\Delta-\frac{i\omega}{2\pi T}\right)}\,,\qquad
    \widehat B(\omega)=\frac{(\pi T)^{\Delta}\,\Gamma\!\left(\frac12-\Delta\right)}{\Gamma\!\left(1-\Delta-\frac{i\omega}{2\pi T}\right)}\,,
\end{equation}
up to a common arbitrary constant factor for the amplitude of the solution.

The third is the response function itself, which is the ratio of the response to the source, with a factor that follows from the normalization \eqref{eq:Onorm},
\begin{align}
    \widehat G_R^{\rm cl}(\omega)&=(2\ell+1)\,\frac{\sqrt\pi\,\Gamma\!\left(\Delta-\frac12\right)}{\Gamma(\Delta)}\,\frac{\widehat B(\omega)}{\widehat A(\omega)}\nonumber\\
    &=\frac{(2\ell+1)\,\pi}{\Gamma(2\Delta)\cos\pi\Delta}\,(2\pi T)^{2\Delta-1}\,\frac{\Gamma\!\left(\Delta-\frac{i\omega}{2\pi T}\right)}{\Gamma\!\left(1-\Delta-\frac{i\omega}{2\pi T}\right)}\,.\label{eq:GRcl}
\end{align}
We omit here, and in the rest of the paper, terms that are even polynomials in $\omega$  with real coefficients. They are local, \ie contact terms at $t=0$, and depend on the renormalization scheme \cite{deHaro:2000vlm,Skenderis:2002wp,Cano:2026scd}; they affect neither the singularities of the response function nor the response at $t>0$.\footnote{In a given black hole the throat response enters the full one with a normalization fixed by the matching to the exterior, which for a minimally coupled scalar in Reissner--Nordstr\"om was obtained in \cite{Cano:2026scd}.} The function \eqref{eq:GRcl} satisfies
\begin{equation}\label{eq:reality}
    \widehat G_R^{\rm cl}(\omega)^*=\widehat G_R^{\rm cl}(-\omega^*)\,,
\end{equation}
as it must for the response to a real perturbation.

\begin{figure}[t]
    \centering
    \includegraphics[width=.9\textwidth]{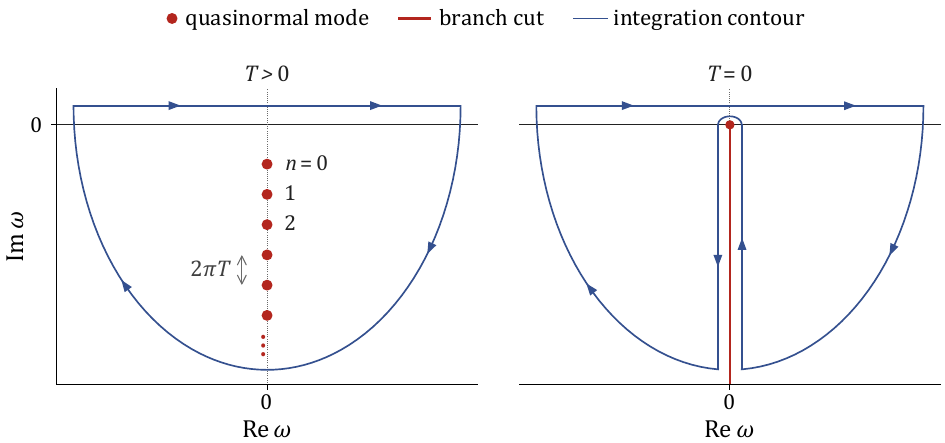}
    \caption{\small Singularities of the classical response function of the throat in the complex frequency plane. Left: at finite temperature the contour encloses the poles at the quasinormal frequencies \eqref{eq:zdm}, with spacing $2\pi T$. Right: at extremality, the poles have condensed into a branch cut from $\omega=0$ along the negative imaginary axis. These are the singularities of the throat alone; the response function of the complete black hole has others at $|\omega|\sim r_0/L_2^2$.}
    \label{fig:classical}
\end{figure}

A quasinormal mode of the throat is a nontrivial solution that is ingoing at the horizon and has no source at the mouth,
\begin{equation}\label{eq:nosource}
    \widehat A(\omega)=0\,.
\end{equation}
When $\Delta$ is not an integer, \eqref{eq:ABcl} shows that this happens at the poles of $\Gamma\!\left(\Delta-\frac{i\omega}{2\pi T}\right)$, which are the frequencies $\omega_n$ in \eqref{eq:zdm}, and there the response function \eqref{eq:GRcl} has poles (the integer case is discussed below). They are its only singularities: at finite temperature the classical response function is meromorphic (figure~\ref{fig:classical}, left).

\paragraph{Beyond the throat.} These are the modes of the throat, and not all the modes of the black hole. In the complete geometry the response at the mouth is matched to the exterior, which shifts the frequencies \eqref{eq:zdm} only at subleading order. In asymptotically flat space part of the perturbation then leaks out as classical radiation, which is the signal seen from outside, but at these low frequencies the leak is suppressed by the greybody factor of the exterior, which brings powers of $\omega r_0$; the modes \eqref{eq:zdm} are then damped by absorption at the horizon.

The complete response function has a richer structure at frequencies $|\omega|\sim r_0/L_2^2$, where the throat description ends. In asymptotically flat space there are, in addition, the damped modes associated with the photon sphere \cite{Zimmerman:2015trm,Berti:2009kk}, and a branch cut along the negative imaginary axis that gives
the Price tail \cite{Price:1971fb}. In asymptotically AdS space the cut is absent, but there are further families of poles \cite{Berti:2009kk}. The modes \eqref{eq:zdm} are the ones that become long-lived near extremality, and the only ones that are significantly modified by the quantum fluctuations of the throat. Those of the photon sphere lie at the limit of the throat description, with $|\omega|\sim1/r_0$, far larger than $E_b$. They probe times $t\sim r_0$, much shorter than $1/E_b$, when the quantum corrections to the response are only of relative order $E_bt$ (section~\ref{subsec:early_late}). For these modes the quantum corrections should then be of relative order $E_br_0=\pi/S_0$.\footnote{Large corrections to these modes were found in \cite{Jiang:2025cyl} by averaging the metric over the Schwarzian fluctuations and extending it outside the throat. But averaging the metric is not averaging the response, and the corrections found there, which are up to order one, are many orders of magnitude larger than the expansion parameter of that analysis.}

\paragraph{Integer $\Delta$.} When the conformal dimension is an integer, as for a massless scalar in Reissner--Nordstr\"om, the response function of the classical throat has no poles. The ratio of Gamma functions in \eqref{eq:GRcl} is then a polynomial in $\omega$, so the throat absorbs, but its response is instantaneous: it is a combination of time derivatives of the source and stops when the source does.\footnote{For half-integer $\Delta$ the polynomial is multiplied by the pole of $1/\cos\pi\Delta$. Once the divergent contact term is subtracted, the poles \eqref{eq:zdm} remain.} Solutions without source and regular on the future horizon still exist at the frequencies \eqref{eq:zdm}, but a source does not excite them: they contain the wave that emerges from the past horizon, which at these frequencies is smooth on the future horizon, while a source at the mouth produces only ingoing waves. They appear instead in two-sided settings, such as the double cone~\cite{Chen:2023hra}.

This absence of poles is peculiar to the classical throat in isolation. The poles return, still classically, when the throat is matched to the exterior,\footnote{As found numerically for gravito-electromagnetic perturbations, which have integer $\Delta$ \cite{Zimmerman:2015trm}.} and also within the throat once its quantum fluctuations are included, as we will see in section~\ref{sec:time_dom_qu_response}.

\subsection{Relaxation in time}
\label{subsec:classical_time}

The response in time is
\begin{equation}\label{eq:GRinv}
    G_R^{\rm cl}(t)=\int_{-\infty+i0}^{+\infty+i0}\frac{d\omega}{2\pi}\,e^{-i\omega t}\,\widehat G_R^{\rm cl}(\omega)\,.
\end{equation}
For $t<0$ the contour closes in the upper half-plane and the response vanishes. For $t>0$ it closes in the lower half-plane and picks the poles \eqref{eq:zdm},\footnote{The arcs at infinity vanish for $0<\Delta<1/2$. For larger $\Delta$ the result follows by analytic continuation in $\Delta$, which only discards contact terms.} with residues
\begin{equation}\label{eq:rescl}
    \mathop{\rm Res}_{\omega=\omega_n}\widehat G_R^{\rm cl}=2i\,(2\ell+1)\sin(\pi\Delta)\,(2\pi T)^{2\Delta}\,\frac{\Gamma(2\Delta+n)}{\Gamma(2\Delta)\,n!}\,.
\end{equation}
The response is then a sum over all the quasinormal modes, which can be done in closed form,
\begin{align}
    G_R^{\rm cl}(t)&=2(2\ell+1)\sin(\pi\Delta)\,(2\pi T)^{2\Delta}\sum_{n=0}^\infty\frac{\Gamma(2\Delta+n)}{\Gamma(2\Delta)\,n!}\,e^{-2\pi T(\Delta+n)t}\nonumber\\
    &=2(2\ell+1)\sin(\pi\Delta)\left[\frac{\pi T}{\sinh\pi Tt}\right]^{2\Delta}.\label{eq:GRclt}
\end{align}
At late times, $t\gg1/T$, only the fundamental mode is left and the decay is exponential, $e^{-2\pi\Delta Tt}$. At early times, $t\ll1/T$, the modes add up to the conformal power law~\cite{Gralla:2016sxp,Zimmerman:2016qtn,Gralla:2018xoz}
\begin{equation}\label{eq:GRearly}
    G_R^{\rm cl}(t)\simeq2(2\ell+1)\sin(\pi\Delta)\,t^{-2\Delta}\,,
\end{equation}
in which the temperature has dropped out, as it should.
This agrees with the commutator \eqref{eq:GRdef} of operators normalized as in \eqref{eq:Onorm},
\begin{equation}\label{eq:commnorm}
    i\left[(it)^{-2\Delta}-(-it)^{-2\Delta}\right]=2\sin(\pi\Delta)\,t^{-2\Delta}\,,
\end{equation}
summed over $m$, which confirms the factor in the first line of \eqref{eq:GRcl}.

\subsection{The extremal limit}
\label{subsec:classical_extremal}

As $T\to0$ the modes \eqref{eq:zdm} become long-lived, and the tower, with spacing $2\pi T$, becomes dense along the negative imaginary axis. At fixed $\omega$ the response function \eqref{eq:GRcl} tends to
\begin{equation}\label{eq:GRext}
    \widehat G_R^{\rm ext}(\omega)=\frac{(2\ell+1)\,\pi}{\Gamma(2\Delta)\cos\pi\Delta}\,(-i\omega)^{2\Delta-1}\,,\qquad -\frac{\pi}{2}<\arg\omega<\frac{3\pi}{2}\,,
\end{equation}
which is no longer meromorphic. It has a branch point at $\omega=0$, and we take the cut along the negative imaginary axis (figure~\ref{fig:classical}, right).\footnote{The cut can be drawn along any curve from $\omega=0$ to infinity in the lower half-plane. The negative imaginary axis is the natural choice: the poles \eqref{eq:zdm} accumulate there as $T\to0$, and it preserves \eqref{eq:reality}.}
The tower has condensed into the cut. Indeed, when $n\gg1$ the residues \eqref{eq:rescl}, divided by the spacing $2\pi T$ between poles, become a smooth function of the position $\zeta=2\pi Tn$ along the axis,
\begin{equation}\label{eq:resdens}
    \frac{1}{2\pi T}\mathop{\rm Res}_{\omega=\omega_n}\widehat G_R^{\rm cl}\simeq \frac{2i\,(2\ell+1)\sin(\pi\Delta)}{\Gamma(2\Delta)}\,\zeta^{2\Delta-1}\,,
    \qquad T \to 0 \,
\end{equation}
so that the sum over poles turns into an integral over $\zeta$ with this weight. The discontinuity across the cut at $\omega=-i\zeta$, i.e., the value on its right minus that on its left, is
then
\begin{equation}\label{eq:disc}
    \mathrm{Disc}\,\widehat G_R^{\rm ext}(-i\zeta)=\frac{4\pi i\,(2\ell+1)\sin(\pi\Delta)}{\Gamma(2\Delta)}\,\zeta^{2\Delta-1}\,.
\end{equation}
When integrating over frequency to go to the time domain, the contour now wraps the cut, and the power law \eqref{eq:GRearly} holds at all times $t>0$.

A single scale, the surface gravity $2\pi T$, controls the classical system: it sets the damping of the modes,
and also the redshift that makes perturbations decay along the horizon.
At extremality both vanish, the throat becomes scale invariant, and the response is left with no scale. This symmetry has a striking consequence on the horizon: transverse derivatives grow unbounded. The phenomenon is the Aretakis instability, to which we return in section~\ref{sec:quantum_Aretakis}.

\subsection{Other throats}
Little in this section depends on the transverse sphere in \eqref{eq:RNAdS}. The index $\ell$ only entered through the value of $\Delta$ and the degeneracy $2\ell+1$. For a throat of the form AdS$_2\times X$ the same analysis applies to any field that reduces to dimension $\Delta$ in AdS$_2$, with $\Delta$ a real number fixed by the spectrum of $X$, and with the sum over $m$ replaced by the sum over the corresponding multiplet of the symmetries of $X$, which only changes the normalization. In the quantum analysis that follows we keep $\Delta$ general, and occasionally refer to $S^2$ as the case of most direct interest.

\section{The perturbed quantum throat}
\label{sec:quantum_setup}

When $E\lesssim E_b$ the geometry \eqref{eq:throat} no longer describes the throat. The description at its mouth survives: a quantum mechanical system, probed by the operators $O_{\Delta\ell m}$ through the coupling \eqref{eq:coupling}, whose response is the retarded Green’s function \eqref{eq:GRdef}. Two things change. First, the soft modes of the throat, whose fluctuations were negligible in the classical black hole, now fluctuate strongly. They can no longer be ignored, and the probe becomes gravitationally dressed. Second, we must specify the state in which the expectation value in \eqref{eq:GRdef} is taken. Classically there was no need: a black hole of given energy and one of given temperature respond in the same way. Now they do not, and a choice must be made. Here we take a state of fixed energy, and leave the state of fixed temperature for \cite{Emparan:2026relaxII}.

\subsection{States of the throat}
\label{subsec:throat_states}

Only a few modes of the throat, described by the Einstein-Maxwell(-AdS) theory, have large fluctuations at low energies. The main one is the Schwarzian mode, which reparametrizes the time at the mouth and has an action proportional to $1/E_b$ \cite{Maldacena:2016upp,Stanford:2017thb,Mertens:2022irh}. The second describes slow rotations of the throat. It is a rotor on $SU(2)$, the symmetry group of $S^2$, with the moment of inertia $I$ of the black hole \cite{Iliesiu:2020qvm}.\footnote{The gauge field of the black hole adds a $U(1)$ mode. Our probes are neutral and the black hole charge is fixed, so it plays no role.} Their Hamiltonian is
\begin{equation}\label{eq:Hthroat}
    H=H_{\rm Sch}+\frac{\mathbf{J}^2}{2I}\,,
\end{equation}
and we label its eigenstates $|E,j,m\rangle$ by the energy $E$ above extremality, the spin $j$ and its projection $m$. The Schwarzian mode has a continuous spectrum, so the states of spin $j$ form a continuum that starts at the rotational energy
\begin{equation}\label{eq:E0j}
    E_{0,j}=\frac{j(j+1)}{2I}\,,
\end{equation}
which is \eqref{eq:E0l} for Reissner--Nordstr\"om. For each $m$, their density is \cite{Stanford:2017thb,Iliesiu:2020qvm}
\begin{equation}\label{eq:rhoj}
    \rho_j(E)=\frac{(2j+1)\,e^{S_0}}{2\pi^2E_b}\,\sinh\!\left(2\pi\sqrt{\frac{2(E-E_{0,j})}{E_b}}\right)\Theta(E-E_{0,j})\,.
\end{equation}
The factor $2j+1$ comes in addition to the degeneracy in $m$, as for a spherical top, whose levels of spin $j$ contain $(2j+1)^2$ states. 
When $E \gg E_b, E_{0,j}$ the logarithm of \eqref{eq:rhoj} is $S_0+4\pi^2T/E_b$, the Bekenstein--Hawking entropy at the temperature \eqref{eq:ET}. At the bottom of each continuum, instead, the density vanishes as $\sqrt{E-E_{0,j}}$. This edge has no classical counterpart, and it will control the response at late times.

The black hole is initially in a state $|E\rangle\equiv|E,0,0\rangle$ of zero spin,\footnote{More precisely, in a microcanonical state: a window of energies of width $\delta E$ around $E$, much narrower than $E_b$, that contains many levels. The window multiplies the survival amplitude \eqref{eq:survival} by its Fourier transform, so our results hold for $t\ll1/\delta E$, which can be exponentially long in $S_0$. Later, the levels in the window dephase.} and a probe of angular momentum $\ell$ takes it to states of spin $\ell$ and energy $E'$. We measure each energy from the bottom of its continuum with the dimensionless momenta (in section~\ref{sec:particle} they are the momenta of a particle)
\begin{equation}\label{eq:kkp}
    k=\sqrt{\frac{2E}{E_b}}\,,\qquad k'=\sqrt{\frac{2(E'-E_{0,\ell})}{E_b}}\,.
\end{equation}
For the massless scalar of \eqref{eq:Delta} in Reissner--Nordstr\"om--AdS$_4$ the rotational energy \eqref{eq:E0j} is
\begin{equation}\label{eq:LambdaAdS}
    E_{0,\ell}=\frac{L^2+r_0^2}{2(L^2+2r_0^2)}\,\Delta(\Delta-1)\,E_b\,,
\end{equation}
which becomes \eqref{eq:E0l} when $L\to\infty$. The semiclassical regime, $E\gg E_b$, is $k\gg1$, and there \eqref{eq:ET} gives \begin{equation}\label{eq:kT}     k=\frac{2\pi T}{E_b}\,. \end{equation} The fully quantum regime is $k\lesssim1$.

\subsection{The probe}
\label{subsec:probe}

The operators $O_{\Delta\ell m}$ carry spin $\ell$, so they connect $|E\rangle$ only to states with $j=\ell$. Their matrix elements between energy eigenstates are known exactly \cite{Mertens:2017mtv,Brown:2024ajk}. Summed over $m$,
\begin{equation}\label{eq:matel}
    \sum_{m=-\ell}^{\ell}\left|\langle E',\ell,m|O_{\Delta\ell m}|E\rangle\right|^2=\frac{2\,e^{-S_0}}{\Gamma(2\Delta)}\left(\frac{E_b}{2}\right)^{2\Delta}\gamma_\Delta(k,k')\,,
\end{equation}
where $\gamma_\Delta$ is the product of four Gamma functions,
\begin{align}
    \gamma_\Delta(k,k')&=\Gamma\big(\Delta+ik+ik'\big)\,\Gamma\big(\Delta+ik-ik'\big)\,\Gamma\big(\Delta-ik+ik'\big)\,\Gamma\big(\Delta-ik-ik'\big)\nonumber\\
    &=\left|\Gamma\big(\Delta+i(k+k')\big)\right|^2\left|\Gamma\big(\Delta+i(k-k')\big)\right|^2\,,\label{eq:gammaDelta}
\end{align}
the second form being valid for real $k$ and $k'$. The normalization is the one fixed in \eqref{eq:Onorm}, as we verify below.
The rotation enters \eqref{eq:matel} only through the threshold $E_{0,\ell}$ in the definition \eqref{eq:kkp} of $k'$: once the spin $\ell$ is fixed, the matrix elements are those of the Schwarzian theory, as functions of $k$ and $k'$.
For real $k$ and $k'$ they are positive and smooth. As functions of complex $k'$ they have poles where the argument of a Gamma function is a non-positive integer, and these are the origin of the poles of the response function that we discuss in section~\ref{sec:quantum_Green}.
These expressions hold for any real $\Delta>0$, integers included.

For a throat AdS$_2\times X$ the rotor moves on the symmetry group of $X$. A probe in a multiplet of this group takes a singlet black hole to states in the same multiplet, whose continuum starts at the corresponding rotational energy. Equations \eqref{eq:rhoj} and \eqref{eq:matel} keep their form, with $2\ell+1$ replaced by the dimension of the multiplet. The space $X$ then enters through two numbers only: the rotational energy $E_{0,\ell}$, and a degeneracy that multiplies the response. So we write our formulas for $S^2$ but treat $\Delta$ and $E_{0,\ell}/E_b$ as independent real parameters. A black hole with initial spin $j$ would instead reach all the spins from $|j-\ell|$ to $j+\ell$, each with its own threshold~\cite{Brown:2024ajk}; we do not consider it explicitly. In parametric estimates
we take $\Delta$ of order one, which is never smaller for the massless fields of section~\ref{sec:classical_Green}. Section~\ref{subsec:stages} discusses large $\Delta$.

\subsection{The perturbed black hole as an unstable state}
\label{subsec:unstable}

A source that acts at $t=0$ puts the black hole in the state $|\psi\rangle=O^\dagger_\Delta|E\rangle$, where we drop the angular indices. It is not an energy eigenstate. By \eqref{eq:matel} it is a superposition of states of spin $\ell$ with all the energies $E'$ of their continuum, and the weight \eqref{eq:weight} of each is the density \eqref{eq:rhoj} times the squared matrix element~\eqref{eq:matel},
\begin{equation}\label{eq:wexplicit}
    \mathsf{W}(E')=\frac{\mathcal{N}}{E_b}\,\sinh(2\pi k')\,\gamma_\Delta(k,k')\,,\qquad E'\geq E_{0,\ell}\,,
\end{equation}
with $k'$ the momentum \eqref{eq:kkp} of the final state, $\mathsf{W}=0$ below $E_{0,\ell}$, and
\begin{equation}\label{eq:Ndef}
    \mathcal{N}=\frac{2\ell+1}{\pi^2\,\Gamma(2\Delta)}\left(\frac{E_b}{2}\right)^{2\Delta}\,.
\end{equation}
The sum over $m$ is included in $\mathsf{W}$, and the factors $e^{\pm S_0}$ have cancelled.

Three properties of $\mathsf{W}$ will be important. First, it has a lower end, at the bottom of the spin-$\ell$ continuum, $E'=E_{0,\ell}$. If $E>E_{0,\ell}$ this lies below $E$: the black hole can give energy to the probe, up to $E-E_{0,\ell}$. If $E<E_{0,\ell}$ it can only absorb, and only quanta with $\omega>E_{0,\ell}-E$ \cite{Emparan:2025qqf,Cano:2026scd}. In both cases, near this end,
\begin{equation}\label{eq:wedge}
    \mathsf{W}(E')\simeq\frac{2\pi\mathcal{N}}{E_b}\,|\Gamma(\Delta+ik)|^4\,k'\,\propto\,\sqrt{E'-E_{0,\ell}}\,.
\end{equation}
Second, at large energies
\begin{equation}\label{eq:wUV}
    \mathsf{W}(E')\to\frac{2\ell+1}{\Gamma(2\Delta)}\,E'^{\,2\Delta-1}\,,
\end{equation}
where $E_b$, $E$ and $E_{0,\ell}$ have dropped out. Third, when $\Delta\geq1/2$ the weight grows monotonically between these two limits (see figure~\ref{fig:weight}).\footnote{From \eqref{eq:wexplicit},
\begin{equation*}
    \frac{d\ln \mathsf{W}}{dk'}=2\pi\coth2\pi k'-2\,\mathrm{Im}\,\psi\big(\Delta+i(k'+k)\big)-2\,\mathrm{Im}\,\psi\big(\Delta+i(k'-k)\big)\,,
\end{equation*}
with $\psi$ the digamma function, and $\mathrm{Im}\,\psi(\Delta+is)\leq\frac{\pi}{2}\tanh\pi s$ for $s>0$ when $\Delta\geq1/2$. For $\Delta<1/2$ the weight falls at large $E'$ as \eqref{eq:wUV}, after a maximum.}

The response follows from the survival amplitude of $|\psi\rangle$,
\begin{equation}\label{eq:aWightman}
    \mathsf{G}(t)=\langle\psi|e^{-i(H-E)t}|\psi\rangle=\sum_{m=-\ell}^{\ell}\langle E|O_{\Delta\ell m}(t)\,O^\dagger_{\Delta\ell m}(0)|E\rangle\,,
\end{equation}
which is the integral \eqref{eq:survival}. It is the first term of the commutator in \eqref{eq:GRdef}, now evaluated in $|E\rangle$. In the usual notation it is the Wightman function $G^>(t)$, and its Fourier transform, as in \eqref{eq:GRFT}, is $\widehat G^>(\omega)=2\pi\,\mathsf{W}(E+\omega)$. The second term, with the opposite ordering, is its complex conjugate, by the reality of the field and the sum over $m$. Then
\begin{equation}\label{eq:GRa}
    G_R(t)=-2\,\Theta(t)\,\mathrm{Im}\,\mathsf{G}(t)\,,
\end{equation}
and its Fourier transform \eqref{eq:GRFT} at real frequencies has
\begin{equation}\label{eq:ImGR}
    \mathrm{Im}\,\widehat G_R(\omega)=\pi\left[\mathsf{W}(E+\omega)-\mathsf{W}(E-\omega)\right]\,.
\end{equation}
The first term is the absorption of a quantum of frequency $\omega$, which takes the black hole to $E+\omega$. The second is the emission stimulated by the source, which takes it to $E-\omega$. Absorption is at least as large as emission at every $\omega>0$, so the black hole never gives net energy to the source.\footnote{For $\Delta\geq1/2$ this follows from the monotonic growth of $\mathsf{W}$. For $\Delta<1/2$, let $k'_a>k'_b\geq0$ be the final momenta of absorption and emission at the same $\omega$. Then $k_a'^2+k_b'^2=2k^2-4E_{0,\ell}/E_b\leq2k^2$, so $|k'_a-k|\leq k-k'_b$, and the factor $|\Gamma(\Delta+i(k-k'))|^2$ of \eqref{eq:gammaDelta} is larger for absorption. The rest, $\sinh(2\pi k')\,|\Gamma(\Delta+i(k+k'))|^2$, has a larger logarithmic derivative than at $\Delta=0$, because $\mathrm{Im}\,\psi(\Delta+is)\leq\mathrm{Im}\,\psi(is)$ for $s>0$, and at $\Delta=0$ its ratio between $k'_a$ and $k'_b$ is at least one, since $k^2\geq k'_ak'_b$.} At low frequency, when $E>E_{0,\ell}$, \eqref{eq:ImGR} becomes $2\pi\omega\,\mathsf{W}'(E)$, and after matching to the exterior this gives the dissipative Love numbers of \cite{Cano:2026scd}.

At short times the integral \eqref{eq:survival} is dominated by large energies, and \eqref{eq:wUV} gives
\begin{equation}\label{eq:aearly}
    \mathsf{G}(t)\to(2\ell+1)\,(it)^{-2\Delta}\,,
\end{equation}
which is \eqref{eq:Onorm} summed over $m$. This confirms the normalization of \eqref{eq:matel}. It also shows that at short enough times (section~\ref{subsec:early_late}) the survival amplitude is the conformal one, whatever the energy of the black hole. For non-integer $\Delta$ so is the response, \eqref{eq:GRearly}, and the quantum effects need time to show up. For integer $\Delta$, \eqref{eq:GRearly} vanishes and the response begins with a quantum effect (section~\ref{subsec:classical_regime}).

At $t=0$, however, the integral diverges, and for $\Delta\geq1/2$ it does not converge at any real~$t$. The state $|\psi\rangle$ is not normalizable, because the source acts at a sharp instant. We regulate it by separating the two insertions of the probe by a short Euclidean time $\epsilon$,
\begin{equation}\label{eq:areg}
    \mathsf{G}(t-i\epsilon)=\int_{E_{0,\ell}}^{\infty}dE'\,\mathsf{W}(E')\,e^{-\epsilon(E'-E)}\,e^{-i(E'-E)t}\,,
\end{equation}
and for $t\neq 0$ the limit $\epsilon\to0$ is finite. The sign of $\epsilon$ damps the unbounded large energies. In $\widehat G_R(\omega)$ the dependence on $\epsilon$ is polynomial in $\omega$. These are the contact terms that we omit, as in section~\ref{sec:classical_Green}; their renormalization is discussed in \cite{Cano:2026scd}.

The regulated state, $e^{-\epsilon(H-E)/2}|\psi\rangle$, has finite squared norm $\mathsf{G}(-i\epsilon)$. The Hamiltonian \eqref{eq:Hthroat} is Hermitian, so the energies in \eqref{eq:areg} are real, and their weights are positive. A sum of phases with positive weights never exceeds its value at $t=0$, so at all real times
\begin{equation}\label{eq:abound}
    |\mathsf{G}(t-i\epsilon)|\leq \mathsf{G}(-i\epsilon)\,,
\end{equation}
and it decays only because its components dephase. The perturbed black hole is unstable only in the way that an unstable quantum state is \cite{Fonda:1978dk}: it decays, and its response is bounded at all times. This holds for any Hermitian Hamiltonian, the classical throat included, so it says nothing about the horizon, where the Aretakis growth takes place. The two features of $\mathsf{W}$ that govern the decay are already in sight: the poles of \eqref{eq:gammaDelta} at complex $k'$, which will give the quasinormal modes, and the edge \eqref{eq:wedge}, which will give the late-time tail.

All of this is the disk approximation, the leading order in $e^{-S_0}$ of the gravitational path integral, in which the Schwarzian and rotational modes are integrated over exactly on the topology of the Euclidean black hole \cite{Stanford:2017thb,Iliesiu:2020qvm,Mertens:2022irh}. The spectrum is then continuous. In the complete theory the black hole has discrete levels, with spacings exponentially small in $S_0$. Then $\mathsf{G}(t)$ is a sum of discrete phases: it stops decaying after times exponentially long in $S_0$, and recurs only much later \cite{Maldacena:2001kr}. The edges of \eqref{eq:rhoj} are resolved at energies exponentially close to them. Our results do not apply to energies $E$ or $E'-E_{0,\ell}$, nor to inverse times, that are exponentially small in $S_0$.

\section{The quantum response at complex frequency}
\label{sec:quantum_Green}

The quasinormal modes are poles of the response function away from the real frequency axis, on the sheets that the function reaches when it is continued across it. In this section we carry out this continuation and determine which of the poles are quasinormal modes.

\subsection{Spectral representation}
\label{subsec:spectral}

The Fourier transform \eqref{eq:GRFT} of the response \eqref{eq:GRa}, with $\mathsf{G}(t)$ given by \eqref{eq:survival}, is
\begin{equation}\label{eq:GRspec}
    \widehat G_R(\omega)=\int_{E_{0,\ell}}^{\infty}dE'\,\mathsf{W}(E')\left[\frac{1}{E'-E-\omega}+\frac{1}{E'-E+\omega}\right],\qquad \mathrm{Im}\,\omega>0\,.
\end{equation}
Here $\omega$ is the frequency of the probe. For $\omega>0$, the first term comes from absorption, and the second from stimulated emission. Only the states with $E'=E+\omega$ in the first term, or $E'=E-\omega$ in the second, are reached by real transitions, and on the real axis, with $\omega\to\omega+i0$, they give \eqref{eq:ImGR}. The others enter virtually. With the growth \eqref{eq:wUV} of the weight, the integral converges only for $0<\Delta<1/2$. For larger $\Delta$ we regulate it as in \eqref{eq:areg}, which multiplies $\mathsf{W}(E')$ by $e^{-\epsilon(E'-E)}$; the terms that diverge when $\epsilon\to0$ are polynomial in $\omega$, and are the contact terms that we omit. Nothing in what follows depends on them.

We measure the energies of the final and the initial state from the bottom of the spin-$\ell$ continuum, in units of $E_b/2$,
\begin{equation}\label{eq:pqdef}
     p=k'^2\,,\qquad q=k^2-2\frac{E_{0,\ell}}{E_b}=\frac{2}{E_b}\big(E-E_{0,\ell}\big)\,,
\end{equation}
so that $p$ is also the squared momentum of the final state. The initial state, and with it $k$ and $q$, is fixed, while the final state ranges over the continuum. Its energy above the initial one is
\begin{equation}
    E'-E=\frac{E_b}{2}\,(p-q)\,.
\end{equation}

As a function of the momentum of the final state, the weight \eqref{eq:wexplicit} is
\begin{equation}\label{eq:FDelta}
    \mathsf{W}(E')=\frac{\mathcal{N}}{E_b}\,\mathsf{F}(k')\,,\qquad \mathsf{F}(u)=\sinh(2\pi u)\,\gamma_\Delta(k,u)\,.
\end{equation}
$\mathsf{W}$ and $\mathsf{F}$ live in different complex planes: in the plane of $E'$, $\mathsf{W}$ has a square-root branch point at the threshold $E' = E_{0,\ell}$,
while in the momentum plane $\mathsf{F}$ is meromorphic. We continue the response in the momentum plane. In terms of $\mathsf{F}$, \eqref{eq:GRspec} becomes
\begin{equation}\label{eq:GRStieltjes}
    \widehat G_R(\omega)=\frac{\mathcal{N}}{E_b}\Big[\mathcal{S}\big(a_+(\omega)\big)+\mathcal{S}\big(a_-(\omega)\big)\Big]\,,
\end{equation}
where, for~$a \in \mathbb{C} \, \backslash \, [0,\infty)$,
\begin{equation}\label{eq:SDelta}
    \mathcal{S}(a)=\int_0^\infty dp\,\frac{\mathsf{F}(\sqrt{p})}{p-a}
\end{equation}
is the Stieltjes transform in the variable $p$ of $\mathsf{F}(\sqrt{p})$, with $\sqrt{p}=k'\geq0$,
and
\begin{equation}\label{eq:apm}
    a_\pm(\omega)=q\pm\frac{2\omega}{E_b}=\frac{2}{E_b}\big(E\pm\omega-E_{0,\ell}\big)\,.
\end{equation}
For real $\omega$, $a_+$ is the squared momentum of the state that the black hole reaches when it absorbs a quantum of frequency $\omega$, and $a_-$ the one it reaches when it emits it. The frequency enters the response only through these two arguments, so its analytic structure follows from that of the single function $\mathcal{S}(a)$.

\subsection{Continuation in the momentum plane}
\label{subsec:continuation}

The integral \eqref{eq:SDelta} defines a function that is analytic in the $a$ plane cut along the positive real axis, where $a$ is the squared momentum of a state of the continuum. We call it the function on the
first sheet, $\mathcal{S}^{\rm I}$. Across the cut it jumps by
\begin{equation}\label{eq:Sdisc}
    \mathcal{S}^{\rm I}(x+i0)-\mathcal{S}^{\rm I}(x-i0)=2\pi i\,\mathsf{F}(\sqrt{x})\,,\qquad x>0\,.
\end{equation}
The cut ends at $a=0$, where the weight vanishes as $\sqrt{p}$, so this is a branch point of square-root type. Around such a point the natural variable is the square root itself, the momentum
\begin{equation}\label{eq:udef}
    u=i\sqrt{-a}\,,
\end{equation}
with the principal square root. This is the root of $u^2=a$ with $\mathrm{Im}\,u>0$. It differs from $\sqrt{a}$, by a sign when $\mathrm{Im}\,a<0$, because its cut lies along the positive real axis, like that of~$\mathcal{S}^{\rm I}$.
The first sheet is therefore the upper half-plane $\mathrm{Im}\,u>0$, and the two sides of the cut are $u(x\pm i0)=\pm\sqrt{x}$.
This $u$ is the momentum $k'$ of the final state, continued to complex values. As a function of complex $u$, $\mathsf{F}$ is odd, because the product \eqref{eq:gammaDelta} is even in its second argument.

In terms of $k'$, with $p=k'^2$ and $a=u^2$, the measure and kernel of \eqref{eq:SDelta} are $dp/(p-a)=dk'\left[1/(k'-u)+1/(k'+u)\right]$.
With $k'\to-k'$ in the second term, the oddness of $\mathsf{F}$ turns \eqref{eq:SDelta} into an integral over the whole real line of momenta,
\begin{equation}\label{eq:SCauchy}
    \mathcal{S}^{\rm I}(a)=\int_{-\infty}^{\infty}dk'\,\frac{\mathsf{F}(k')}{k'-u}\,,\qquad \mathrm{Im}\,u>0\,.
\end{equation}
The continuation is now elementary. The integrand of \eqref{eq:SCauchy} depends on $u$ only through the kernel, and the pole of the kernel at $k'=u$ is the only singularity that can reach the contour: the poles of $\mathsf{F}$ are fixed at $k'=\pm k\pm i(\Delta+n)$, at a distance of at least $\Delta$ from the real line.
When $u$ crosses the real axis into the lower half-plane, the contour must be deformed to stay below the pole, which adds its residue.
The integral over the real line that remains is~\eqref{eq:SCauchy} evaluated at $-u$, as the change $k'\to-k'$ shows, that is, $\mathcal{S}^{\rm I}(u^2)$.
The analytic continuation of the Stieltjes transform is given through the function
\begin{equation}\label{eq:Scont}
    \widetilde{\mathcal{S}}(u)=
    \begin{cases} \mathcal{S}^{\rm I}(u^2) \,,\quad & \mathrm{Im}\,u>0\,, \\
    \mathcal{S}^{\rm I}(u^2)+2\pi i\,\mathsf{F}(u)\,,\quad & \mathrm{Im}\,u<0\,,
    \end{cases}
\end{equation}
and on the real axis by the common limit of the two, which agree by~\eqref{eq:Sdisc}.

The function $\widetilde{\mathcal S}(u)$ is then single valued and meromorphic in the whole plane. Its upper half is the first sheet of the $a$ plane and its lower half is the second, and there are no others. The only singularities are the poles of $\mathsf{F}$ in the lower half-plane. The point of the second sheet that lies over $a$ is $-u(a)$, so in terms of $a$ \eqref{eq:Scont} reads
\begin{equation}\label{eq:SII}
    \mathcal{S}^{\rm II}(a)=\mathcal{S}^{\rm I}(a)-2\pi i\,\mathsf{F}\big(u(a)\big)\,,
\end{equation}
with $u(a)$ the root \eqref{eq:udef}.
This holds on the whole cut plane, whichever side of the cut is crossed.Equivalently, \eqref{eq:SII} follows in one step from \eqref{eq:Sdisc},
by the edge-of-the-wedge theorem or Morera's theorem
because the jump is the value on the cut of a function that is analytic near it.
The regulator multiplies $\mathsf{F}(k')$ by an entire function of $k'$, which changes nothing in the argument and drops out of \eqref{eq:Scont} and \eqref{eq:SII} when $\epsilon\to0$.

We now return to the frequency. The two terms of \eqref{eq:GRStieltjes} have momenta
\begin{equation}\label{eq:upm}
    u_\pm(\omega)=i\sqrt{-a_\pm(\omega)}=i\sqrt{\frac{2}{E_b}\big(E_{0,\ell}-E\mp\omega\big)}\,,
\end{equation}
and each has its branch point where $a_\pm=0$, at the threshold frequencies
\begin{equation}\label{eq:thresholds}
    \omega=\pm\big(E_{0,\ell}-E\big)\,.
\end{equation}
The cuts lie on the real axis,
\begin{equation}\label{eq:cuts}
    \mathcal{C}_+=\big[E_{0,\ell}-E,\,+\infty\big)\,,\qquad \mathcal{C}_-=\big(-\infty,\,E-E_{0,\ell}\big]\,.
\end{equation}
$\mathcal{C}_+$ contains the frequencies that the black hole can absorb and $\mathcal{C}_-$ those that it can emit: they are the supports of the two terms of \eqref{eq:ImGR}. If $E<E_{0,\ell}$ the cuts leave a gap $|\omega|<E_{0,\ell}-E$ in which the response is real, and the black hole neither absorbs nor emits \cite{Emparan:2025qqf,Cano:2026scd}. If $E>E_{0,\ell}$ they overlap on $|\omega|<E-E_{0,\ell}$.

Near a threshold the two sheets differ only in the sign of an odd function of $u$. Because $\mathsf{F}$ is odd, the function
\begin{equation}\label{eq:Rdef}
    \mathcal{R}(a)=\widetilde{\mathcal{S}}(u)-i\pi\,\mathsf{F}(u)
\end{equation}
is even in $u$, and therefore a single-valued function of $a=u^2$. It has no cut. Its only singularities are the poles of $\mathsf{F}$, at $a=\big(k\pm i(\Delta+n)\big)^2$, so it is analytic in the disk $|a|<k^2+\Delta^2$. Appendix~\ref{app:Stieltjes} gives it as a sum over these poles. On the two sheets, with $u(a)$ the root \eqref{eq:udef},
\begin{equation}\label{eq:SIR}
    \mathcal{S}^{\rm I}(a)=\mathcal{R}(a)+i\pi\,\mathsf{F}\big(u(a)\big)\,,\qquad
    \mathcal{S}^{\rm II}(a)=\mathcal{R}(a)-i\pi\,\mathsf{F}\big(u(a)\big)\,.
\end{equation}
All that is not analytic at $a=0$ is in the term with $\mathsf{F}$, which changes sign from one sheet to the other. Near $u=0$ it is $\mathsf{F}(u)\simeq2\pi|\Gamma(\Delta+ik)|^4\,u$, as in \eqref{eq:wedge}. When $E\neq E_{0,\ell}$ the other transform in \eqref{eq:GRStieltjes} has no branch point at the same frequency. So near each threshold \eqref{eq:thresholds}, on the first sheet,
\begin{equation}\label{eq:GRthreshold}
    \widehat G_R(\omega)\simeq\text{analytic}+\frac{2\pi^2 i\,\mathcal{N}}{E_b}\,\big|\Gamma(\Delta+ik)\big|^4\,u_\pm(\omega)\,,
\end{equation}
for $|u_\pm|\ll1$, i.e. $|\omega\mp(E_{0,\ell}-E)|\ll E_b$.

Each of the two transforms in \eqref{eq:GRStieltjes} can be on its first or on its second sheet, so the response function has four sheets. The response function is therefore meromorphic on a four-sheeted Riemann surface over the frequency plane, with square-root branch points at the two thresholds. We label them by a pair of numerals, the first for the sheet of the emission term and the second for that of the absorption term, in the order of their cuts $\mathcal{C}_-$ and $\mathcal{C}_+$ along the real axis. The retarded function in the upper half-plane is on $(\mathrm{I},\mathrm{I})$. Crossing $\mathcal{C}_+$ changes the second label, crossing $\mathcal{C}_-$ the first, and crossing their overlap changes both. From \eqref{eq:SII},
\begin{align}
    \widehat G_R^{(\mathrm{I},\mathrm{II})}(\omega)&=\widehat G_R^{(\mathrm{I},\mathrm{I})}(\omega)-\frac{2\pi i\,\mathcal{N}}{E_b}\,\mathsf{F}\big(u_+(\omega)\big)\,,\nonumber\\
    \widehat G_R^{(\mathrm{II},\mathrm{I})}(\omega)&=\widehat G_R^{(\mathrm{I},\mathrm{I})}(\omega)-\frac{2\pi i\,\mathcal{N}}{E_b}\,\mathsf{F}\big(u_-(\omega)\big)\,,\nonumber\\
    \widehat G_R^{(\mathrm{II},\mathrm{II})}(\omega)&=\widehat G_R^{(\mathrm{I},\mathrm{I})}(\omega)-\frac{2\pi i\,\mathcal{N}}{E_b}\Big[\mathsf{F}\big(u_+(\omega)\big)+\mathsf{F}\big(u_-(\omega)\big)\Big]\,,\label{eq:GRsheets}
\end{align}
where $\widehat G_R^{(\mathrm{I},\mathrm{I})}$ is \eqref{eq:GRStieltjes} with both transforms given by the integral \eqref{eq:SDelta}, and $u_\pm$ are the roots \eqref{eq:upm}.\footnote{When $E=E_{0,\ell}$ the two branch points coincide at $\omega=0$, and the retarded function reaches only half of each sheet: the upper halves of $(\mathrm{I},\mathrm{I})$ and $(\mathrm{II},\mathrm{II})$, and the lower halves of $(\mathrm{I},\mathrm{II})$ and $(\mathrm{II},\mathrm{I})$.}

The sheet $(\mathrm{I},\mathrm{I})$ is analytic away from the real axis and has no poles. Its lower half
holds the advanced response through the relation~$\widehat G_R^{(\mathrm{I},\mathrm{I})}(\omega^*)=\widehat G_R^{(\mathrm{I},\mathrm{I})}(\omega)^*$.
The poles are on the other three sheets.

\subsection{Poles and quasinormal modes}
\label{subsec:poles}

The poles of \eqref{eq:GRsheets} are those of $\mathsf{F}$.\footnote{For the BTZ black hole at large central charge, the quasinormal modes also come from the poles of the matrix elements of a probe between heavy states \cite{Turiaci:2016cvo}. The matrix elements \eqref{eq:matel} are their Schwarzian limit \cite{Mertens:2017mtv}. The real parts of the BTZ modes come from the momentum along the circle, and are classical.} On the second sheet, the lower half of the $u$ plane, they come from the two Gamma functions $\Gamma(\Delta\pm ik-iu)$ in \eqref{eq:gammaDelta}, and lie at
\begin{equation}\label{eq:upoles}
    u=\pm k-i(\Delta+n)\,,\qquad n=0,1,2,\dots
\end{equation}
The momentum of the final state is that of the initial one displaced by $-i(\Delta+n)$. This lowers the real part of its energy by the recoil $(\Delta+n)^2E_b/2$, which we interpret in section~\ref{subsec:particle_recoil}. The frequencies follow from \eqref{eq:apm}, which in the two channels gives
\begin{equation}\label{eq:omegau}
    \omega=\pm\frac{E_b}{2}\left(u^2-q\right)\,,
\end{equation}
with the upper sign for absorption and the lower one for emission. One pole of each channel then lies in the lower half of the frequency plane,
\begin{align}
    \text{absorption:}\qquad u&=k-i(\Delta+n)\,, & \omega&=\omega_n^+\,,\nonumber\\
    \text{emission:}\qquad u&=-k-i(\Delta+n)\,, & \omega&=\omega_n^-\,,\label{eq:polemap}
\end{align}
at the frequencies \eqref{eq:qqnm}. They are related by
\begin{equation}\label{eq:polereality}
    \omega_n^-=-\left(\omega_n^+\right)^*\,,
\end{equation}
as the reality condition \eqref{eq:reality} requires. The other pole of each channel gives the complex conjugate frequency, in the upper half-plane. It does not conflict with causality, because it is not on $(\mathrm{I},\mathrm{I})$. The sheet $(\mathrm{I},\mathrm{II})$ thus has the poles $\omega_n^+$ and their conjugates, $(\mathrm{II},\mathrm{I})$ has $\omega_n^-$ and their conjugates, and $(\mathrm{II},\mathrm{II})$ has them all.

Not all these poles are quasinormal modes. The response in time is the integral \eqref{eq:GRinv} of the retarded function along the real axis, taken from above. The exponentials appear when the contour is lowered into the lower half-plane, and lowering it continues the function across the real axis, at each frequency, straight down. A pole is met only if it lies on this continuation: on the sheet that is reached from the point of the real axis directly above it. This is how a resonance appears in scattering theory: as a pole of the sheet adjacent to the physical region \cite{Fonda:1978dk}.

The same condition can be read off the plane when the two cuts are rotated to run straight down from the branch points \eqref{eq:thresholds}, as in figure~\ref{fig:pruning}. Moving a cut does not change the function, only the sheet that each region of the plane displays. With the cuts on the real axis the lower half-plane displays $(\mathrm{I},\mathrm{I})$ and no poles; with vertical cuts it displays, below each point of the real axis, the continuation from that point, and the contour can be lowered freely until it wraps the two cuts and the poles in view.

\begin{figure}[t]
    \centering
    \includegraphics[width=.55\textwidth]{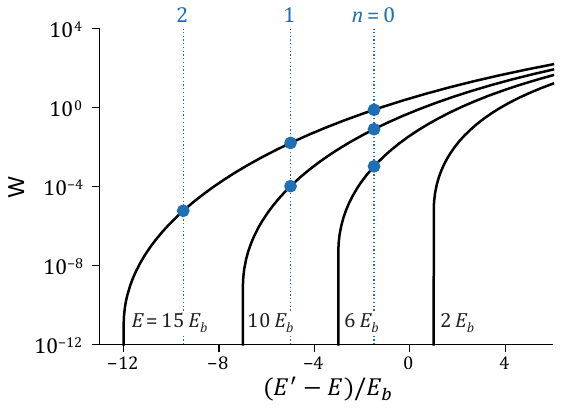}
    \caption{\small The weight $\mathsf{W}(E')$, in units of $(2\ell+1)E_b^{2\Delta-1}$, as a function of $E'-E$, for the example of figure~\ref{fig:pruning} at the same four energies. Each curve starts at the bottom of the spin-$\ell$ continuum, $E'=E_{0,\ell}$. The dotted lines mark $E'-E=\mathrm{Re}\,\omega_n^+$, the energies of the poles \eqref{eq:qqnm} above $E$, which do not depend on $E$. A pole is a quasinormal mode when its line crosses the curve (dots), and it is pruned when the line lies to the left of the curve, where the weight vanishes. The weight has no peaks at the dots, so these are not sharp resonances.}
    \label{fig:weight}
\end{figure}

Below a frequency that lies on $\mathcal{C}_+$ the absorption term is on its second sheet, and elsewhere it stays on the first, where it has no poles. The pole $\omega_n^+$ is therefore a quasinormal mode if and only if
\begin{equation}\label{eq:QNMcond}
    \mathrm{Re}\,\omega_n^+>E_{0,\ell}-E\qquad\Longleftrightarrow\qquad k>\Delta+n\,,
\end{equation}
which is the condition \eqref{eq:prune}. The same holds for $\omega_n^-$ and $\mathcal{C}_-$. In words, the energy $E+\mathrm{Re}\,\omega_n^+$ of the resonance must lie within the continuum of spin-$\ell$ states.\footnote{The pole $\omega_n^-$ gives the same energy, at $\omega=-\mathrm{Re}\,\omega_n^+$, so the two poles of a pair describe a single resonance of the black hole. A source of positive frequency reaches it by absorption, through $\omega_n^+$, if it lies above $E$, and by stimulated emission, through $\omega_n^-$, if it lies below, \ie if the recoil $(\Delta+n)^2E_b/2$ exceeds $E_{0,\ell}$. This is the case for every massless scalar, by \eqref{eq:Lambdabound}, but not for the fundamental pair of the gravito-electromagnetic perturbations with $\Delta=\ell$~\cite{Zimmerman:2015trm}.} Figure~\ref{fig:weight} shows this on the real axis. The energies $\mathrm{Re}\,\omega_n^+$ do not depend on $E$, while the lower end of the weight moves up as $E$ is lowered and leaves them behind one by one. The weight passes smoothly through those that remain. For $\Delta\geq1/2$ it grows monotonically (section~\ref{subsec:unstable}), so the resonances leave no peak in it.

\begin{figure}[t]
    \centering
    \includegraphics[width=.9\textwidth]{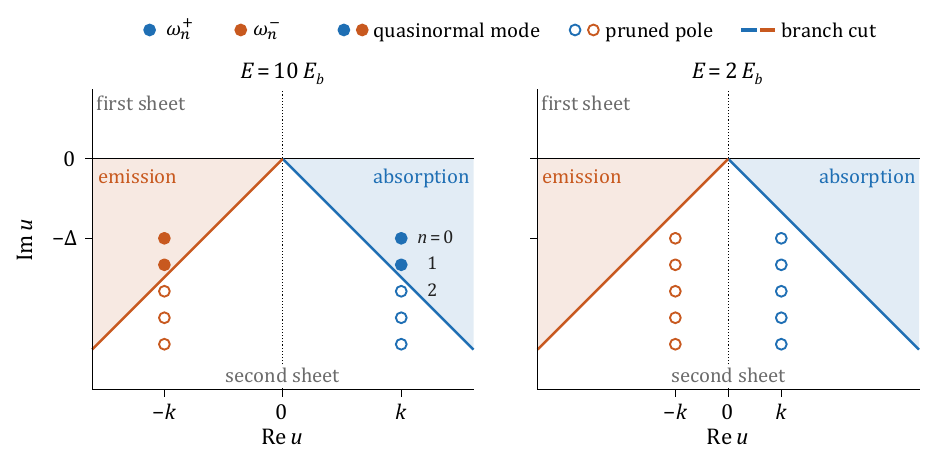}
    \caption{\small The momentum plane of the final state, for the example of figure~\ref{fig:pruning} at two energies. The upper half is the first sheet and the lower half the second. The poles lie at $u=\pm k-i(\Delta+n)$. Continuing the retarded function straight down from the real frequency axis reaches the shaded sectors, the right one for absorption and the left one for emission; their edges are the vertical cuts of figure~\ref{fig:pruning}. Poles inside the sectors (filled) are quasinormal modes; the others (hollow) are pruned. As $E$ is lowered the poles move horizontally towards the imaginary axis and leave the sectors.}
    \label{fig:kplane}
\end{figure}

The momentum plane makes this transparent (figure~\ref{fig:kplane}). Going down from a point of $\mathcal{C}_+$ at fixed $\mathrm{Re}\,\omega$ keeps $\mathrm{Re}\,u^2$ positive, so $u$ leaves the positive real axis into the sector
\begin{equation}\label{eq:sector}
    -\frac{\pi}{4}<\arg u<0\,.
\end{equation}
Going down from a point to the left of the threshold, $u$ never leaves the upper half-plane. The part of the lower half $\omega$ plane to the right of the vertical cut is then the sector \eqref{eq:sector}, and the cut is its edge, $\arg u=-\pi/4$. The absorption pole in \eqref{eq:polemap} is inside the sector when the real part of its momentum exceeds the imaginary part,
\begin{equation}\label{eq:insector}
    \mathrm{Re}\,u>|\mathrm{Im}\,u|\qquad\Longleftrightarrow\qquad k>\Delta+n\,,
\end{equation}
which is \eqref{eq:QNMcond}. For emission the sector is the mirror image,
\begin{equation}\label{eq:sectorem}
    -\pi<\arg u<-\frac{3\pi}{4}\,,
\end{equation}
and the emission pole in \eqref{eq:polemap} is inside it under the same condition.

As the energy is lowered the poles \eqref{eq:upoles} move horizontally towards the imaginary axis, and they leave the sectors one by one, the deepest first. The quasinormal modes are the pairs $\omega_n^\pm$ with
\begin{equation}\label{eq:nmodes}
    0\leq n<k-\Delta\,.
\end{equation}
Their damping satisfies
\begin{equation}\label{eq:dampbound}
    \left|\mathrm{Im}\,\omega_n^\pm\right|=E_bk\,(\Delta+n)>(\Delta+n)^2E_b\geq\Delta^2E_b\,,
\end{equation}
so none is damped more slowly than $\Delta^2E_b$, and when $k\leq\Delta$, \ie $E\leq\Delta^2E_b/2$, none is left. A pruned pole does not disappear: it sits behind the cut, on a sheet that the retarded function reaches only by going around the branch point. Its frequency remains complex. When $k=\Delta+n$ the pole crosses the vertical cut at
\begin{equation}\label{eq:polecross}
    \omega_n^+=E_{0,\ell}-E-i(\Delta+n)^2E_b\,,\qquad u=(\Delta+n)(1-i)\,,
\end{equation}
far below the branch point, and it goes on behind the cut with a damping that is still proportional to $\sqrt{E}$. A resonance in potential scattering behaves differently: it reaches the imaginary momentum axis and turns there into a bound or a virtual state of real energy. In our case the poles come from the matrix elements \eqref{eq:matel} of the probe, not from the spectrum. Their momenta \eqref{eq:upoles} have real parts $\pm k$, those of the initial state, so they reach the imaginary axis, at real frequency, only in the formal limit $E\to0$ (section~\ref{subsec:limits}).

The vertical cuts are singled out among all the possible ones because $e^{-i\omega t}$ decays along them without oscillating, so the integral around them contains no exponentials. In section~\ref{sec:time_dom_qu_response} we show that the response in time is exactly this integral plus the modes that satisfy \eqref{eq:QNMcond}.

\subsection{The classical regime and the low-energy limit}
\label{subsec:limits}

\paragraph{Classical regime.}
When $E\gg E_b$ the momentum $k$ is large, and the state has the temperature in \eqref{eq:ET}, related to $k$ by \eqref{eq:kT}.
Consider frequencies of order $T$, which are much smaller than $E$. The thresholds \eqref{eq:thresholds} are far away,
\begin{equation}\label{eq:thresholdscl}
    \omega=\pm\big(E_{0,\ell}-E\big)\simeq\mp E\,,
\end{equation}
so these frequencies lie on the overlap of the two cuts, and the continuation from above lands on $(\mathrm{II},\mathrm{II})$. There, the poles \eqref{eq:qqnm} are
\begin{equation}\label{eq:polescl}
    \omega_n^\pm=-2\pi iT(\Delta+n)\pm\left(E_{0,\ell}-\frac{E_b}{2}(\Delta+n)^2\right)\,.
\end{equation}
For $n\ll k$ the real part, of order $(\Delta+n)^2E_b$, is negligible compared with the damping, and the two poles of each pair merge into the classical mode \eqref{eq:zdm}. The pruned poles have $n>k-\Delta$, and they are damped faster than
\begin{equation}\label{eq:prunedcl}
    \left|\mathrm{Im}\,\omega_n^\pm\right|>2\pi Tk=2E\,.
\end{equation}

The function itself becomes the classical one. At these frequencies the state reached by absorption, $E'=E+\omega$, has momentum
\begin{equation}\label{eq:kpcl}
    k'\simeq k+\frac{\omega}{2\pi T}\,,
\end{equation}
and the weight tends to
\begin{equation}\label{eq:wthermal}
    \mathsf{W}(E+\omega)\to\frac{2\ell+1}{2\pi\,\Gamma(2\Delta)}\,(2\pi T)^{2\Delta-1}\,e^{\omega/2T}\left|\Gamma\!\left(\Delta+\frac{i\omega}{2\pi T}\right)\right|^2\,.
\end{equation}
It satisfies detailed balance,
\begin{equation}\label{eq:detbal}
    \mathsf{W}(E-\omega)=e^{-\omega/T}\,\mathsf{W}(E+\omega)\,,
\end{equation}
so at these frequencies the state of fixed energy cannot be told apart from the thermal one. Moreover, in \eqref{eq:ImGR} it gives
\begin{align}
    \mathrm{Im}\,\widehat G_R(\omega)&=\pi\left[\mathsf{W}(E+\omega)-\mathsf{W}(E-\omega)\right]\nonumber\\
    &\to\frac{2\ell+1}{\Gamma(2\Delta)}\,(2\pi T)^{2\Delta-1}\,\sinh\!\left(\frac{\omega}{2T}\right)\left|\Gamma\!\left(\Delta+\frac{i\omega}{2\pi T}\right)\right|^2\,,\label{eq:ImGRcl}
\end{align}
which is the imaginary part of the classical response \eqref{eq:GRcl}. Then \eqref{eq:GRspec} reproduces this response in the upper half-plane, up to contact terms. Its lower half-plane, with the poles \eqref{eq:zdm}, is the limit of the sheet $(\mathrm{II},\mathrm{II})$, while the lower half of $(\mathrm{I},\mathrm{I})$ tends to the classical advanced function, $\widehat G_A^{\rm cl}(\omega)=\widehat G_R^{\rm cl}(\omega^*)^*=\widehat G_R^{\rm cl}(-\omega)$, which is a different function:
\begin{align}
    \widehat G_R^{(\mathrm{I},\mathrm{I})}(\omega)&\to\widehat G_R^{\rm cl}(\omega)\,, & \mathrm{Im}\,\omega&>0\,,\nonumber\\
    \widehat G_R^{(\mathrm{II},\mathrm{II})}(\omega)&\to\widehat G_R^{\rm cl}(\omega)\,, & \mathrm{Im}\,\omega&<0\,,\nonumber\\
    \widehat G_R^{(\mathrm{I},\mathrm{I})}(\omega)&\to\widehat G_A^{\rm cl}(\omega)\,, & \mathrm{Im}\,\omega&<0\,.\label{eq:sheetscl}
\end{align}
The classical response is meromorphic because the branch points have receded to infinity, leaving $(\mathrm{I},\mathrm{I})$ above and $(\mathrm{II},\mathrm{II})$ below joined along the whole real axis. Along it, $(\mathrm{I},\mathrm{I})$ below differs from $(\mathrm{I},\mathrm{I})$ above by $2i\,\mathrm{Im}\,\widehat G_R$, so
the advanced function is not the continuation of the retarded one.

The corrections to \eqref{eq:wthermal} are of relative order $E_{0,\ell}/T$ and $\Delta^2E_b/T$. The classical regime thus requires not only $E\gg E_b$ but also
\begin{equation}\label{eq:clcond}
    T\gg E_{0,\ell}\,,\qquad T\gg\Delta^2E_b\,,
\end{equation}
i.e., a temperature that is large compared with the rotational energy and with the recoil energy of the probe. For the massless scalar of \eqref{eq:Delta}, \eqref{eq:LambdaAdS} gives
\begin{equation}\label{eq:Lambdabound}
    E_{0,\ell}\leq\frac{\Delta(\Delta-1)}{2}\,E_b\,,
\end{equation}
and the first condition in \eqref{eq:clcond} follows from the second. The condition $E\gg E_b$ then suffices, although the corrections decrease slowly, as $\sqrt{E_b/E}$.

\paragraph{Low-energy limit.}
Now lower the energy at fixed $E_b$. Once $E\lesssim E_b$ the state has no temperature to speak of, and we describe it by its energy only. The thresholds \eqref{eq:thresholds} stay at finite frequency,
\begin{equation}\label{eq:thresholdslow}
    \omega=\pm\big(E_{0,\ell}-E\big)\to\pm E_{0,\ell}\,.
\end{equation}
The poles approach the real axis, with damping proportional to $\sqrt{E}$ and fixed real parts,
\begin{equation}\label{eq:poleslow}
    \omega_n^\pm\to\pm\left(E_{0,\ell}-\frac{E_b}{2}(\Delta+n)^2\right)\,,
\end{equation}
but each of them is pruned on the way, when its damping has come down to $(\Delta+n)^2E_b$. They reach the axis to the left of the threshold of $\mathcal{C}_+$ (to the right of that of $\mathcal{C}_-$) and behind the cut, where the retarded function is smooth. In the momentum plane they end at
\begin{equation}\label{eq:upoleslow}
    u=\pm k-i(\Delta+n)\to-i(\Delta+n)\,,
\end{equation}
at a distance $\Delta+n$ from the real momenta. Nothing condenses.

What remains near the real axis are the two thresholds, with the square roots \eqref{eq:GRthreshold}. If $E_{0,\ell}>0$ they stay apart, separated by the gap. If $E_{0,\ell}=0$ they meet at $\omega=0$ when $E\to0$, and the two terms of \eqref{eq:GRthreshold} add up to
\begin{equation}\label{eq:GRzero}
    \widehat G_R(\omega)\simeq\text{regular}-\frac{4\pi^2\mathcal{N}}{E_b}\,\Gamma(\Delta)^4\left(-\frac{i\omega}{E_b}\right)^{1/2}\,, \qquad \mathrm{Im}\,\omega>0\,,\quad E\ll|\omega|\ll E_b\,.
\end{equation}
This is a branch point at zero frequency, but it is not the conformal one of \eqref{eq:GRext}: its exponent does not depend on $\Delta$, and it comes from the edge of the density of states, not from scale invariance.

The classical extremal limit takes $E_b\to0$ first, which sends the thresholds to infinity and merges the pairs, and then $T\to0$, which makes the tower condense into the branch point of \eqref{eq:GRext}. At fixed $E_b$ this never happens, so the limits $E_b\to0$ and $\omega\to0$ do not commute. Even in the quantum case the strict limit $E\to0$ is formal: energies exponentially small in~$S_0$ lie beyond the disk approximation (section~\ref{sec:quantum_setup}).

\section{Time domain response and stages of the relaxation}
\label{sec:time_dom_qu_response}

We now obtain the response in time. It splits exactly into the integral around the vertical cuts of section~\ref{sec:quantum_Green} and the quasinormal modes, and from this the stages of the relaxation follow.

\subsection{The exact decomposition}
\label{subsec:decomposition}

By \eqref{eq:GRa} the response is $-2\,\mathrm{Im}\,\mathsf{G}(t)$, so we work with the survival amplitude. In terms of the momentum $k'$ of the final state, with \eqref{eq:pqdef} and \eqref{eq:FDelta}, the integral \eqref{eq:survival} is
\begin{equation}\label{eq:Akprime}
    \mathsf{G}(t)=\mathcal{N}\,e^{-i(E_{0,\ell}-E)t}\int_0^\infty dk'\,k'\,\mathsf{F}(k')\,e^{-iE_bk'^2t/2}\,.
\end{equation}
We rotate the contour to the ray $k'=e^{-i\pi/4}s$, with $s>0$, closing it with an arc at infinity.
This ray is the edge of the sector \eqref{eq:sector}, \ie the vertical cut that runs down from the threshold of $\mathcal{C}_+$, on which
\begin{equation}\label{eq:zetadef}
    \omega=E_{0,\ell}-E-i\zeta\,,\qquad \zeta=\frac{E_b}{2}\,s^2\,,
\end{equation}
and the oscillating factor in \eqref{eq:Akprime} becomes $e^{-\zeta t}$. With the regulator of \eqref{eq:areg} the arc does not contribute when $t>0$, and on the ray the regulator can be removed. The rotation sweeps the sector, and in it $\mathsf{F}$ has the poles $u=k-i(\Delta+n)$ with $\Delta+n<k$, as in \eqref{eq:insector}. Then
\begin{equation}\label{eq:Adecomp}
    \mathsf{G}(t)=\mathsf{G}^{\rm cut}(t)-i\sum_{\Delta+n<k}r_n^+\,e^{-i\omega_n^+t}\,,
\end{equation}
where
\begin{equation}\label{eq:Acut}
    \mathsf{G}^{\rm cut}(t)=-\frac{i\,\mathcal{N}}{E_b}\,e^{-i(E_{0,\ell}-E)t}\int_0^\infty d\zeta\,e^{-\zeta t}\,\mathsf{F}\!\left(e^{-i\pi/4}\sqrt{\frac{2\zeta}{E_b}}\right)
\end{equation}
and
\begin{align}
    r_n^+&=2\pi\mathcal{N}\,u\mathop{\rm Res}_{u=k-i(\Delta+n)}\mathsf{F}(u)\nonumber\\
    &=-2\pi^2\mathcal{N}\,\big(k-i(\Delta+n)\big)\,\frac{\sinh2\pi(k-i\Delta)}{\sinh2\pi k}\,\frac{\Gamma(2\Delta+n)\,\Gamma(2\Delta+n+2ik)}{n!\,\Gamma(n+1+2ik)}\,.\label{eq:rn}
\end{align}
The response is therefore
\begin{equation}\label{eq:GRdecomp}
    G_R(t)=G_R^{\rm cut}(t)+\sum_{\Delta+n<k}\left(r_n^+\,e^{-i\omega_n^+t}+r_n^-\,e^{-i\omega_n^-t}\right)\,,\qquad r_n^-=\left(r_n^+\right)^*\,,
\end{equation}
with
\begin{equation}\label{eq:Gcut}
    G_R^{\rm cut}(t)=-2\,\mathrm{Im}\,\mathsf{G}^{\rm cut}(t)\,.
\end{equation}

This is the lowered contour of section~\ref{subsec:poles}. By \eqref{eq:GRsheets}, $\mathsf{F}$ on the ray is the discontinuity of the response function across the vertical cut of the absorption channel, so $\mathsf{G}^{\rm cut}$ is the integral around that cut. The cut of the emission channel and the poles $\omega_n^-$ give the complex conjugate.

The decomposition \eqref{eq:GRdecomp} is exact for all $t>0$, for any energy and any $\Delta>0$, integers included. Away from the crossing values $k=\Delta+n$, the integral over $\zeta$ converges absolutely without a regulator.\footnote{At a crossing a pole lies on the integration ray, and the full cut-plus-pole expression is defined by continuity in $E$. The limits from either side agree.} Contact terms never appear. The sum is finite, with the modes \eqref{eq:nmodes}, and it contains only the poles that satisfy \eqref{eq:QNMcond}: the exponentials in the response are those of the quasinormal modes and no others.\footnote{When $E_{0,\ell}=(\Delta+n)^2E_b/2$ the two poles of the $n$-th pair coincide, and their terms combine into $2\,\mathrm{Re}\,r_n^+\,e^{-i\omega_n^+t}$, which vanishes at isolated values of $E$.} Classically, instead, the response \eqref{eq:GRclt} is a sum over the infinite tower of modes, which at early times add up to the conformal decay; here that decay comes from the cut (section~\ref{subsec:early_late}). When $k\leq\Delta$ the sum is empty and
\begin{equation}\label{eq:GRonlycut}
    G_R(t)=G_R^{\rm cut}(t)\,.
\end{equation}

The pruning is sharp for the poles but smooth in the response. When $k$ decreases through $\Delta+n$ the pole crosses the ray, and its term leaves the sum and is absorbed into the integral.\footnote{At large $t$ this is a Stokes phenomenon: at the crossing the term is of order $e^{-(\Delta+n)^2E_bt}$, exponentially smaller than the integral, which decays as a power.}

\subsection{Early and late times}
\label{subsec:early_late}

The two ends of the relaxation are power laws, and both come from $G_R^{\rm cut}$: the sum in \eqref{eq:GRdecomp} is finite, so it is regular when $t\to0$ and exponentially small when $t\to\infty$.

\paragraph{Early times.}
When $t\to0$ the integral \eqref{eq:Acut} is dominated by large $\zeta$, \ie by the large energies of \eqref{eq:wUV}, and gives the conformal decay \eqref{eq:aearly}. The first correction to it is
\begin{equation}\label{eq:Aearly2}
    \mathsf{G}(t)=(2\ell+1)\,(it)^{-2\Delta}\left[1-iE_\star t+\mathcal{O}(t^2)\right]\,,\qquad E_\star=E_{0,\ell}-\frac{\Delta(\Delta-1)}{6}\,E_b\,,
\end{equation}
so that
\begin{equation}\label{eq:GRearly2}
    G_R(t)=2(2\ell+1)\,t^{-2\Delta}\left[\sin\pi\Delta+E_\star t\,\cos\pi\Delta+\mathcal{O}(t^2)\right]\,.
\end{equation}
The energy of the black hole does not appear at this order. The expansion holds while the corrections are small, which requires $t\ll1/E_b$, $t\ll1/E_{0,\ell}$ and $t\ll1/\sqrt{E_bE}$. Until then a black hole of any energy responds like the extremal classical throat \eqref{eq:GRearly}, except near integer~$\Delta$: there $\sin\pi\Delta$ is small, and at $t\sim|\tan\pi\Delta|/|E_\star|$
the response passes to~\eqref{eq:GRearlyint} for integer $\Delta$.

\paragraph{Late times.}
When $t\to\infty$ the integral \eqref{eq:Acut} is dominated by small $\zeta$, the bottom of the continuum, where
\begin{equation}\label{eq:Fsmall}
    \mathsf{F}(u)\simeq2\pi\,|\Gamma(\Delta+ik)|^4\,u\,,
\end{equation}
as in \eqref{eq:wedge}. Then
\begin{align}
    \mathsf{G}^{\rm cut}(t)&\to\frac{\pi^{3/2}}{2}\,\mathcal{N}\,|\Gamma(\Delta+ik)|^4\left(\frac{E_bt}{2}\right)^{-3/2}e^{-i\left[(E_{0,\ell}-E)t+3\pi/4\right]}\,,\nonumber\\
    G_R^{\rm cut}(t)&\to\pi^{3/2}\,\mathcal{N}\,|\Gamma(\Delta+ik)|^4\left(\frac{E_bt}{2}\right)^{-3/2}\sin\!\left[(E_{0,\ell}-E)\,t+\frac{3\pi}{4}\right]\,,\label{eq:tail}
\end{align}
for $t\gg1/E_b$. The phase converges slowly: its first correction is of order $2\pi^2/(E_bt)$, so the oscillating factor is accurate only once $E_bt$ reaches a few hundred,
while the envelope $t^{-3/2}$ sets in much earlier (figure~\ref{fig:stages}).
The exponent does not depend on $\Delta$: it is fixed by the square root of the density of states \eqref{eq:rhoj} at its edge.
The tail oscillates at the threshold frequency $|E-E_{0,\ell}|$, which is not the frequency of any mode, and it does not oscillate when $E=E_{0,\ell}$. The modes decay exponentially, so at late enough times the tail dominates the response of a black hole of any energy. This is the late-time departure from exponential decay of an unstable quantum system whose spectrum is bounded from below \cite{Khalfin:1958,Fonda:1978dk}. Like the edge itself, the tail is a result of the disk approximation, and it does not extend to times that are exponentially long in $S_0$. In asymptotically flat space, Hawking emission ends it earlier, although only after times longer than $1/E_b$ by powers of $S_0$ (section~\ref{sec:concl}).

\subsection{The classical regime}
\label{subsec:classical_regime}

When $E\gg E_b$ the momentum $k$ is large and the temperature is given by \eqref{eq:kT}. For $n\ll k$ the residues \eqref{eq:rn} become
\begin{equation}\label{eq:rncl}
    r_n^\pm\to\pm i\,e^{\mp i(\pi\Delta+\delta_n)}\,(2\ell+1)\,(2\pi T)^{2\Delta}\,\frac{\Gamma(2\Delta+n)}{\Gamma(2\Delta)\,n!}\,,
\end{equation}
with a small phase that we keep,
\begin{equation}\label{eq:deltan}
    \delta_n=\frac{(2\Delta+1)(\Delta+n)}{2k}=\frac{(2\Delta+1)(\Delta+n)\,E_b}{4\pi T}\,.
\end{equation}
Each pole of a pair carries half of the classical residue \eqref{eq:rescl} divided by $\sin\pi\Delta$, and the pair contributes to the response
\begin{equation}\label{eq:paircl}
    r_n^+\,e^{-i\omega_n^+t}+r_n^-\,e^{-i\omega_n^-t}\to2(2\ell+1)\,(2\pi T)^{2\Delta}\,\frac{\Gamma(2\Delta+n)}{\Gamma(2\Delta)\,n!}\,e^{-2\pi T(\Delta+n)t}\,\sin\!\big(\pi\Delta+\delta_n+\mathrm{Re}\,\omega_n^+\,t\big)\,.
\end{equation}
The factor $\sin\pi\Delta$ of the classical mode has become the interference between the two poles of the pair. The phases $\delta_n$ and $\mathrm{Re}\,\omega_n^+\,t$ are small for the modes that matter when
\begin{equation}\label{eq:clwindow}
    \frac{1}{E}\ll t\ll\frac{1}{E_b}\,,\ \frac{1}{E_{0,\ell}}\,,
\end{equation}
and then \eqref{eq:paircl} is the $n$-th term of the classical sum \eqref{eq:GRclt}. The quantum response is the classical one in this window, with corrections of relative order $\Delta^2E_b/T$ and $|\mathrm{Re}\,\omega_n^+|\,t$, multiplied by $|\cot\pi\Delta|$ near integer $\Delta$. Before it, at $t\lesssim1/E$, the overtones with $n\gtrsim k$ are needed. They are pruned, and their effect is in $G_R^{\rm cut}$, but there the decay is the conformal one \eqref{eq:GRearly2} in any case.

The window \eqref{eq:clwindow} ends when the phases $\mathrm{Re}\,\omega_n^+t$ in \eqref{eq:paircl} become of order one. The exponential decay goes on, now with a slow change of sign, until the tail \eqref{eq:tail} emerges. For large $k$ the amplitude of the tail is exponentially small,
\begin{equation}\label{eq:tailamp}
    |\Gamma(\Delta+ik)|^4\simeq4\pi^2\,k^{4\Delta-2}\,e^{-2\pi k}\,,
\end{equation}
because the bottom of the continuum lies far below the energies that the probe excites. The fundamental pair decays as $e^{-\Delta E_bk\,t}$, and it falls below the tail at
\begin{equation}\label{eq:tcross}
    t_*\simeq\frac{2\pi}{\Delta E_b}\,,
\end{equation}
up to logarithmic corrections. This time does not depend on the energy of the black hole. The exponential stage lasts from $t\sim1/T$ to $t_*$, and in it the response falls by a factor of order $e^{-2\pi k}$. Figure~\ref{fig:stages} (left) shows it for $k=10$ in the modulus of the survival amplitude, where the classical result follows the exact one until it falls through the tail.

\paragraph{Integer $\Delta$.}
Nothing in \eqref{eq:GRdecomp} is special when $\Delta$ is an integer: the poles, the residues \eqref{eq:rn} and the pruning are the same. What changes is the interference. In \eqref{eq:paircl} the two poles of a pair now cancel at leading order, and leave
\begin{equation}\label{eq:pairint}
    r_n^+\,e^{-i\omega_n^+t}+r_n^-\,e^{-i\omega_n^-t}\to2(-1)^\Delta(2\ell+1)\,(2\pi T)^{2\Delta}\,\frac{\Gamma(2\Delta+n)}{\Gamma(2\Delta)\,n!}\,e^{-2\pi T(\Delta+n)t}\,\sin\!\big(\delta_n+\mathrm{Re}\,\omega_n^+\,t\big)\,,
\end{equation}
which is smaller than for generic $\Delta$ by a factor of order $E_b/T$ or $E_bt$, and vanishes in the classical limit. This is how the poles that the classical response with integer $\Delta$ lacks (section~\ref{subsec:classical_QNMs}) return in the quantum throat: each pair has the classical damping, and its contribution to the response is a quantum effect. At early times the first term of \eqref{eq:GRearly2} vanishes for the same reason, the leading term of \eqref{eq:Aearly2} being real, and the second gives
\begin{equation}\label{eq:GRearlyint}
    G_R(t)\simeq2(-1)^\Delta(2\ell+1)\,E_\star\,t^{1-2\Delta}\,,
\end{equation}
instead of contact terms only. For a massless scalar in Reissner--Nordstr\"om, \eqref{eq:E0l} gives
\begin{equation}\label{eq:EstarRN}
    E_\star=\frac{\Delta(\Delta-1)}{3}\,E_b\,.
\end{equation}
The survival amplitude still decays as $t^{-2\Delta}$, and the late-time tail \eqref{eq:tail} is unchanged.\footnote{For $\ell=0$, where $\Delta=1$ and $E_\star=0$, the response has no singular early stage and begins linearly in $t$.}

\subsection{Stages and ringing}
\label{subsec:stages}

The stages of the relaxation follow from \eqref{eq:GRdecomp} and its limits.\footnote{We consider generic $\Delta$; the modifications for integer $\Delta$ are given in section~\ref{subsec:classical_regime}.}

When $E\gg E_b$ there are three stages. The response decays as $t^{-2\Delta}$ for $t\ll1/T$, then exponentially, as the classical sum \eqref{eq:GRclt} and at the end as its fundamental mode, until the time $t_*$ of \eqref{eq:tcross}, and from then on as the tail \eqref{eq:tail}.

When $E\leq\Delta^2E_b/2$ there are no quasinormal modes, the response is \eqref{eq:GRonlycut}, and there are two stages: the conformal decay \eqref{eq:GRearly2} for $t\ll1/E_b$ and the tail \eqref{eq:tail} for $t\gg1/E_b$, with no exponential stage in between. This is the right panel of figure~\ref{fig:stages}. For $\Delta>3/4$ the passage is not monotonic: around $t\sim1/E_b$ the response falls faster than the conformal law before it joins the tail, which decays more slowly.

At intermediate energies a few modes remain, whose decay times, by \eqref{eq:dampbound}, are shorter than $1/(\Delta^2E_b)$ and comparable to the time $1/E_b$ at which the tail sets in. They show only as a short exponential transient between the two power laws.

For large $\Delta$, as at high angular momentum, $\Delta^2E_b$ is much larger than $E_b$. The exponential stage then needs $E\gg\Delta^2E_b$, and by \eqref{eq:nmodes} a black hole with $E\gg E_b$ may have no quasinormal modes at all. The conformal decay also ends earlier than $1/E_b$, when the corrections in \eqref{eq:Aearly2}, which grow as $\Delta^2E_bt$, become large.

\paragraph{Ringing.}
The modes have real parts, but they hardly ring. For a mode that satisfies \eqref{eq:QNMcond}, the ratio of the frequency to the damping is bounded,
\begin{equation}\label{eq:quality}
    \frac{|\mathrm{Re}\,\omega_n^\pm|}{|\mathrm{Im}\,\omega_n^\pm|}=\frac{\big|E_{0,\ell}-(\Delta+n)^2E_b/2\big|}{k\,(\Delta+n)\,E_b}\leq\frac{\big|E_{0,\ell}-(\Delta+n)^2E_b/2\big|}{(\Delta+n)^2E_b}\,.
\end{equation}
For the massless scalar of \eqref{eq:Delta}, the bound \eqref{eq:Lambdabound} makes the right-hand side at most $1/2$: the phase of every mode turns by less than half a radian in each $e$-folding of its decay. A mode could complete oscillations only if the rotational energy were much larger than the recoil,
\begin{equation}\label{eq:ringcond}
    E_{0,\ell}\gg\frac{(\Delta+n)^2}{2}\,E_b\,,
\end{equation}
which the sphere by itself does not provide. When $E\gg E_b$, over the whole duration of the exponential stage the phase of the fundamental pair turns by
\begin{equation}
    |\mathrm{Re}\,\omega_0^+|\,t_*\approx\frac{2\pi}{\Delta E_b}\left|\frac{\Delta^2E_b}{2}-E_{0,\ell}\right|\,.
\end{equation}
For the massless scalar in Reissner--Nordstr\"om this is approximately $\pi$,
so the response changes sign about once before the tail, and the only sustained oscillation is that of the late-time tail, at the threshold frequency $|E-E_{0,\ell}|$.

The two power laws are separated by the scale $E_b$, however small the energy. When $E_b\to0$ first, the conformal decay lasts forever at fixed $E$, and this is the classical extremal throat of section~\ref{subsec:classical_extremal}.
At fixed $E_b$ it ends at $t\sim1/E_b$.
We examine the consequences of this fact for the Aretakis instability in section~\ref{sec:quantum_Aretakis}.

\section{A particle picture of the relaxation}
\label{sec:particle}

\begin{figure}[t]
\centering
\includegraphics[width=\textwidth]{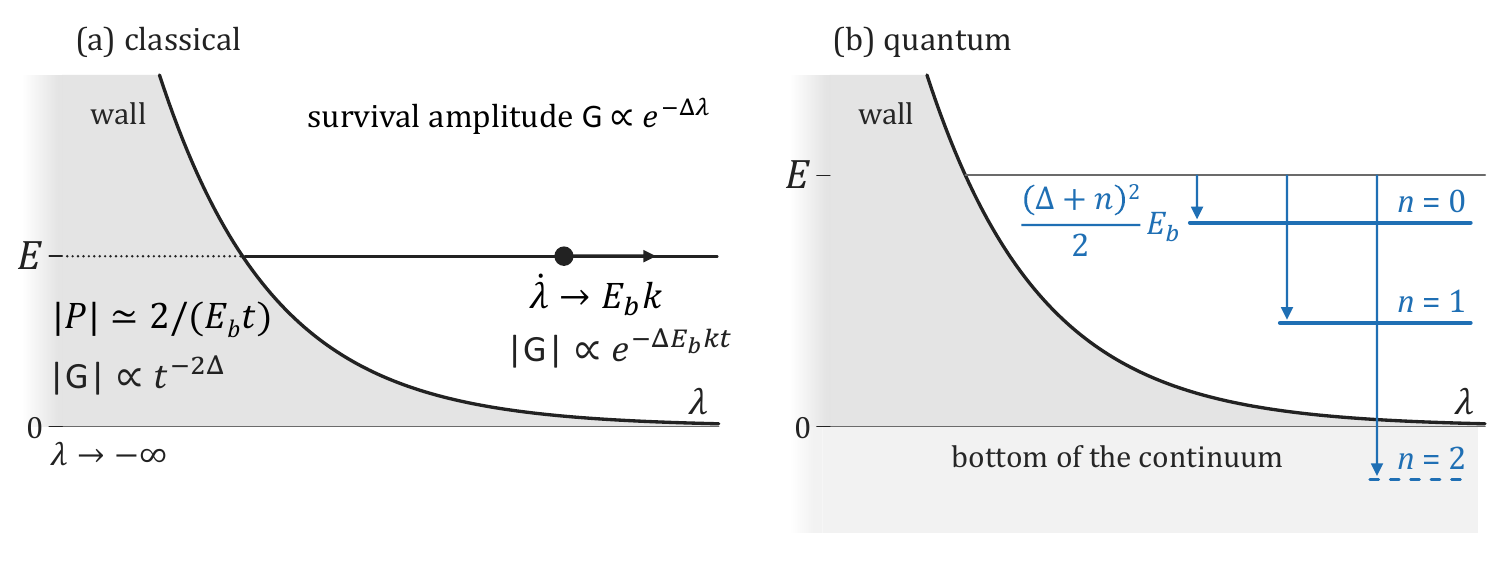}
    \caption{\small The particle picture. The renormalized length $\lambda$ of a geodesic through the throat behaves as a particle of mass $1/E_b$ in the potential $2E_be^{-\lambda}$. The survival amplitude is proportional to $e^{-\Delta\lambda}$, so it decays as the particle moves away from the wall, exponentially once $\lambda$ grows at a constant velocity far from it. (a) Classically, the geodesic of the response starts at zero length, \ie at $\lambda\to-\infty$, deep inside the wall (dotted), where a particle of energy $E$ cannot be. Its length is complex, and the figure shows its real part (appendix~\ref{app:particle}). While inside the wall it has momentum $|P|\simeq2/(E_bt)$, which gives the conformal decay. At $t\sim1/T$ it comes out of the wall and moves on with velocity $\dot\lambda\to E_bk$, which gives the damping $e^{-\Delta E_bk\,t}$ of the fundamental mode. (b) Quantum mechanically, the probe changes the momentum of the particle by $-i(\Delta+n)$ for the overtone $n$, and the recoil lowers the energy of the excited state by $(\Delta+n)^2E_b/2$, shown here for $\ell=0$. The lines are the real parts of the energies of the modes, not levels of the particle. The modes whose energy remains above the bottom of the continuum are quasinormal modes (solid); the others are pruned (dashed).}
\label{fig:particle}
\end{figure}

The relaxation also has a simple picture in terms of the Schwarzian mode of the throat. In JT gravity, the Hilbert space of a two-sided black hole is that of a single collective coordinate, the length $\lambda$ of the geodesic that joins the two boundaries, in units of the AdS$_2$ radius and renormalized by subtracting its divergence at the boundaries \cite{Harlow:2018tqv}. Its dynamics is Liouville quantum mechanics \cite{Bagrets:2016cdf,Harlow:2018tqv,Kitaev:2018wpr}. The Hamiltonian $H_{\rm Sch}$ of \eqref{eq:Hthroat}, which advances the time of either boundary towards the future, is
\begin{equation}\label{eq:Liouville}
H_{\rm Sch}=\frac{E_b}{2}\left(P^2+4e^{-\lambda}\right),
\end{equation}
where $P$ is the momentum conjugate to $\lambda$. This is a particle of mass $1/E_b$ that moves freely at large $\lambda$ and is reflected by an exponential wall at small $\lambda$ (figure~\ref{fig:particle}). A black hole of energy $E$ above extremality is a stationary state of the particle: a standing wave with momenta $\pm k$ far from the wall, where
\begin{equation}\label{eq:Ek}
E=\frac{E_b}{2}\,k^2\,.
\end{equation}
In states of spin $\ell$ the rotation adds the energy $E_{0,\ell}$.

The same quantum-mechanical system describes the response.
The survival amplitude at time $t$ is the amplitude for a perturbation that enters the throat at time $0$ to come back out at time $t$,
leaving the black hole in its initial state. In between, the perturbation travels along the geodesic through the throat that joins the two times, and picks up a factor $e^{-\Delta\lambda}$ from its renormalized length $\lambda$ \cite{Yang:2018gdb,Mertens:2022irh}, so
\begin{equation}\label{eq:alambda}
\mathsf{G}(t)\sim e^{-\Delta\lambda}\,.
\end{equation}
This geodesic has both ends on the same boundary, but its length is still the coordinate $\lambda$ of \eqref{eq:Liouville}. The geodesic splits the disk of the Euclidean gravitational path integral into two pieces, each bounded by the geodesic and by one of the two arcs into which the insertions divide the boundary. Each piece prepares a state of a two-sided black hole whose slice is the geodesic, so the path integral on it is a wave function of $\lambda$ for the Hamiltonian \eqref{eq:Liouville}.
Here we give the picture in simple terms; appendix~\ref{app:particle} makes it precise, and rederives from it the quasinormal frequencies and the stages of the decay.

\subsection{Classical motion}
\label{subsec:particle_classical}

The particle moves classically while its wavelength is short compared with the scale of the wall, i.e., $|P|\gg1$, and while the probe hardly changes its motion.
Both hold when $E_b$ is small at fixed temperature, so that $k=2\pi T/E_b\gg\Delta^2$,
and at times short compared with $1/E_b$. The trajectory then gives the relaxation
of the classical black hole, in two stages.

At time $0$ the two ends of the geodesic coincide, so its renormalized length starts at $\lambda\to-\infty$, deep inside the wall.
No real trajectory of energy $E$ starts there, and the one that does is complex, as we now show. Figure~\ref{fig:particle}a shows its real part.\footnote{The imaginary part of $\lambda$ stays equal to $\pi$ at all times (appendix~\ref{app:particle}), so the real part moves in the inverted potential. This is why the particle slows down as it comes out.}

At early times, $t\lesssim1/T$, the particle is deep inside the wall. The energy $E$ is then negligible compared with the potential, and the particle moves as in the extremal throat, with zero energy,
\begin{equation}
    P^2+4e^{-\lambda}=0\,.
\end{equation}
For real $\lambda$ this is impossible: the kinetic term cannot cancel a positive potential. With $\dot\lambda=E_bP$ it gives $\frac{d}{dt}e^{\lambda/2}=\pm iE_b$, so a geodesic that has zero length at $t=0$ has
\begin{equation}
    e^{\lambda/2}\simeq iE_bt\,,\qquad\text{i.e.}\qquad \lambda\simeq2\ln(E_bt)+i\pi\,,\qquad |P|\simeq\frac{2}{E_bt}\,,
\end{equation}
with the sign of the continuation from Euclidean time, $t=-i\tau$, as in \eqref{eq:Onorm}. The length acquires the imaginary part $\pi$, which inverts the potential, and its real part grows logarithmically. Then
\begin{equation}
    E_b^{2\Delta}\,e^{-\Delta\lambda}\simeq(it)^{-2\Delta}\,,
\end{equation}
which is the conformal decay \eqref{eq:Onorm}, phase included. This holds while the potential is much larger than $E$, that is, until $4e^{-\mathrm{Re}\,\lambda}\simeq4/(E_bt)^2$ falls to $k^2$, where the particle leaves the wall, at $t\simeq2/(E_bk)=1/(\pi T)$.

After that, at any nonzero temperature, the particle is far from the wall. Its energy is then almost all kinetic, so its velocity approaches a constant, and the survival amplitude decays exponentially,
\begin{equation}\label{eq:classdecay}
\dot\lambda\to E_bk\,,\qquad \mathsf{G}(t)\sim e^{-\Delta E_bk\,t}\,.
\end{equation}
With $E_bk=2\pi T$, this is the damping of the fundamental mode.

The overtones come from the force of the wall, which is proportional to $e^{-\lambda}$. Far from the wall the particle moves almost uniformly, with $\mathrm{Re}\,\lambda\simeq E_bkt$, so the force decays as $e^{-E_bkt}$ and adds to $\lambda$ a correction proportional to $e^{-E_bkt}$. The force depends on the position of the particle, so this correction changes it by a relative amount of order $e^{-E_bkt}$, which adds to $\lambda$ a correction proportional to $e^{-2E_bkt}$, and so on. The survival amplitude is then a series
\begin{equation}\label{eq:overtones}
\mathsf{G}(t)\sim e^{-\Delta E_bk\,t}\left(1+c_1\,e^{-E_bk\,t}+c_2\,e^{-2E_bk\,t}+\dots\right),
\end{equation}
whose $n$-th term decays at the rate $(\Delta+n)E_bk$. For $E\gg E_b$ these are the overtones \eqref{eq:zdm}. The rates are equally spaced because the wall is exactly exponential in $\lambda$.

The classical description fails when the wavelength of the particle, $1/|P|$, grows to the distance over which the wall varies, a change of order one in $\lambda$. Inside the wall $|P|\simeq2/(E_bt)$, which falls to order one at $t\sim1/E_b$. When $E\lesssim E_b$ this happens before the particle leaves the wall. The scale invariance of the early stage is then broken by $E_b$ rather than by the energy, and quantum effects end the conformal decay at $t\sim1/E_b$, before any exponential decay has had time to develop (section~\ref{subsec:stages}).

When $E\gg E_b$ the particle leaves the wall long before, at $t\sim1/T$, and its momentum never falls below $k$, so its wavelength stays short. Even so, the motion \eqref{eq:classdecay} holds only until $t\sim1/E_b$. Then the recoil of the particle, which we discuss next, shifts the phase of the modes, and later the tail \eqref{eq:tailparticle} takes over.

\subsection{Quantum recoil and pruning}
\label{subsec:particle_recoil}

Quantum mechanically, the particle recoils when it absorbs the probe: this is the backreaction of the probe on the state of the particle. The probe multiplies the wave function of the particle by $e^{-\Delta\lambda}$, which can be seen as a plane wave with imaginary momentum. Far from the wall the black hole state is made of the waves $e^{\pm ik\lambda}$. Counting momenta towards the wall, $k$ labels the ingoing wave $e^{-ik\lambda}$, which the probe turns into $e^{-i(k+\delta k)\lambda}$, with $\delta k$ imaginary. The outgoing wave gives no further modes: in stimulated emission it gives the partners $\omega_n^-$. Including the overtones, which come from the wall, the change in momentum is, as appendix~\ref{app:particle} shows,
\begin{equation}\label{eq:q}
\delta k=-i(\Delta+n)\,,
\end{equation}
and the energy of the particle changes by
\begin{equation}\label{eq:dE}
\frac{E_b}{2}(k+\delta k)^2-\frac{E_b}{2}\,k^2=-i(\Delta+n)E_bk-\frac{(\Delta+n)^2}{2}\,E_b\,.
\end{equation}
The two terms are those of an atom of mass $1/E_b$ and velocity $E_bk$ that absorbs a photon of momentum $\delta k$: a Doppler shift, $E_bk\,\delta k$, and a recoil shift, $E_b(\delta k)^2/2$. Adding the rotational energy $E_{0,\ell}$ to \eqref{eq:dE} gives the frequency $\omega_n^+$ of \eqref{eq:qqnm}.

For the imaginary $\delta k$ of \eqref{eq:q} the Doppler shift is imaginary, and it gives
the damping:
the profile $e^{-(\Delta+n)\lambda}$ is carried towards the wall at velocity $E_bk$, so at any fixed $\lambda$ it decays in time at the classical rate. This is the amplitude of the classical motion read from the other end of the geodesic. There we followed the particle forward in time from zero length; here we follow, also forward in time, the state that the probe produces, and read it at zero length. The recoil shift is real and negative. It lowers the frequency by $(\Delta+n)^2E_b/2$ below $E_{0,\ell}$, so the mode oscillates. A classical particle is infinitely heavy, so at fixed velocity it has no recoil, and its modes are purely damped.

For integer $\Delta$, as for the massless scalar in Reissner--Nordstr\"om, the classical response \eqref{eq:GRclt} vanishes, since $\sin\pi\Delta=0$. The classical motion then gives only the envelope of the survival amplitude, and the response comes entirely from quantum effects: the decay $t^{1-2\Delta}$ of \eqref{eq:GRearlyint} at early times, and later the quantum phases of the modes in \eqref{eq:pairint}.

The recoil lowers the energy of the excited state. A mode is a resonance only if this energy, $E+E_{0,\ell}-(\Delta+n)^2E_b/2$, remains within the continuum of spin-$\ell$ states, which starts at $E_{0,\ell}$ (section~\ref{subsec:poles}), i.e., only while $E$ exceeds the recoil. This is the pruning condition \eqref{eq:prune} (figure~\ref{fig:particle}b). Without recoil, no mode is pruned.

\subsection{Late-time tail}
\label{subsec:particle_tail}

The excited state is a superposition of states of the particle with momenta $k'$, whose phases $e^{-iE_bk'^2t/2}$ drift apart in time. At late times only the slowest ones, with $k'\lesssim(E_bt)^{-1/2}$, still add up. These particles barely move, and the wall reflects them as a hard wall would. Their weight per unit $k'$ vanishes as $k'^2$, which is the edge \eqref{eq:wedge} of the weight, so, with $t\to t-i\epsilon$ as in \eqref{eq:areg},
\begin{align}\label{eq:tailparticle}
\mathsf{G}(t)&\propto e^{-i(E_{0,\ell}-E)t}\int_0^\infty dk'\,k'^2\,e^{-iE_bk'^2t/2}\nonumber\\
&\propto e^{-i(E_{0,\ell}-E)t}\,(E_bt)^{-3/2}\,,
\end{align}
as for a free particle on a half-line. The tail oscillates at the frequency $|E-E_{0,\ell}|$ of the transition to the bottom of the continuum. When $E\gg E_b$ the weight of these slow states is of order $e^{-2\pi k}$, by \eqref{eq:tailamp}, and they overtake the exponentials only at the time $t_*$ of \eqref{eq:tcross}.
The response thus decays first as $t^{-2\Delta}$, then exponentially if $E\gg E_b$, and finally as $t^{-3/2}$.

\section{Aretakis in a quantum throat}
\label{sec:quantum_Aretakis}

Away from extremality, a perturbation of a black hole decays on the future horizon together with all its derivatives \cite{Kay:1987ax,Dafermos:2008ys}. This is due to the redshift, which grows exponentially in the advanced time on the horizon at a rate set by the surface gravity $2\pi T$, the same rate that damps the modes \eqref{eq:zdm}. At extremality the surface gravity vanishes and the redshift is lost. Outside the horizon the perturbation still decays together with all its derivatives, and on the horizon the perturbation itself decays, but its transverse derivatives of high enough order grow without bound. This is the Aretakis instability.

We pose the problem as a response, in the same way as we treated the relaxation in section~\ref{sec:classical_Green}. A source $A_{\ell m}(t)$ acts at the mouth of the throat during a finite time, and we follow the field that it produces as it propagates through the throat and reaches the horizon. Once the source is off, this field provides the data for the subsequent evolution on any slice that crosses the horizon. These data are not the most general ones, because they are generated by a single function at the mouth, but for non-integer $\Delta$ they are enough to make the transverse derivatives grow.
This formulation suits the quantum problem: the Schwarzian theory gives us the response at the mouth, and in the classical throat the field on the horizon follows from it.\footnote{Nonlinear evolutions also suggest that the growth belongs to the throat more than to the horizon: they find it also where no horizon forms, and where one does, before any trapped surface appears \cite{Gelles:2026bix}.}

In this formulation the growth on the horizon is tied to the analytic structure of section~\ref{sec:classical_Green}. We will see that the growth and the decay \eqref{eq:GRearly} at the mouth of the extremal throat are two limits of a single solution, whose form is fixed by the scaling symmetry of AdS$_2$~\cite{Gralla:2018xzo}.
The condensing tower of modes describes how the near-extremal throat approaches this solution. Quantum effects break the symmetry, and in section~\ref{subsec:Aretakis_qu} we examine what is left of the instability.

To follow the field onto the horizon we use the advanced time
\begin{equation}\label{eq:vEF}
    v=t+\frac{1}{4\pi T}\ln\frac{R}{R+4\pi T}\,,
\end{equation}
which becomes $v=t-1/R$ when $T=0$, and in which the metric \eqref{eq:throat} is
\begin{equation}\label{eq:EF}
    ds^2=L_2^2\left[-R(R+4\pi T)\,dv^2+2\,dv\,dR\right]+r_0^2\,d\Omega^2\,,
\end{equation}
regular on the future horizon $R=0$. Points of the horizon have finite $v$ and are reached as $t\to\infty$; at the mouth, $v\to t$. A field of dimension $\Delta$ satisfies
\begin{equation}\label{eq:KGEF}
    2\,\partial_v\partial_R\Psi_{\ell m}+\partial_R\big[R(R+4\pi T)\,\partial_R\Psi_{\ell m}\big]-\Delta(\Delta-1)\,\Psi_{\ell m}=0\,,
\end{equation}
and taking $N-1$ derivatives with respect to $R$ at $R=0$ gives the evolution of the $N$-th transverse derivative along the horizon,
\begin{equation}\label{eq:horizonid}
    \partial_v\,\partial_R^N\Psi_{\ell m}=-2\pi TN\,\partial_R^N\Psi_{\ell m}+\frac{(\Delta-N)(\Delta+N-1)}{2}\,\partial_R^{N-1}\Psi_{\ell m}\,,\qquad R=0\,.
\end{equation}
The first term is the redshift, which damps the $N$-th derivative at the rate $2\pi TN$. When $T=0$ it is absent, and each derivative is driven by the one before: if $\partial_R^{N-1}\Psi_{\ell m}$ falls as $v^{N-1-\Delta}$, then $\partial_R^N\Psi_{\ell m}$ goes as $v^{N-\Delta}$, and it grows when $N>\Delta$. When $\Delta$ is an integer the last term vanishes for $N=\Delta$, and $\partial_R^\Delta\Psi_{\ell m}$ is conserved along the horizon: this is the Aretakis constant \cite{Aretakis:2012ei,Lucietti:2012xr}.

\subsection{The extremal throat}
\label{subsec:Aretakis_ext}

\paragraph{Self-similar growth.}
When $T=0$ the metric \eqref{eq:EF} is invariant under the scalings $v\to\alpha v$, $R\to R/\alpha$, with $\alpha>0$. A source $A_{\ell m}(t)=\delta(t)$ at the mouth produces the retarded solution $\mathcal{K}(v,R)$, the bulk-to-boundary propagator of the throat, and the scaling symmetry requires $\mathcal{K}(\alpha v,R/\alpha)=\alpha^{-\Delta}\mathcal{K}(v,R)$. In the coordinates \eqref{eq:vEF}, and normalized to a unit source, it is~\cite{Gralla:2018xzo}
\begin{equation}\label{eq:Kext}
    \mathcal{K}(v,R)=\frac{2\,\Gamma(\Delta)\sin\pi\Delta}{\sqrt{\pi}\,\Gamma\!\left(\Delta-\frac12\right)}\,\frac{\Theta(v)}{\big[v\,(2+Rv)\big]^{\Delta}}\,.
\end{equation}
For $v>0$ it solves \eqref{eq:KGEF}, it is regular on the horizon, and at $R\to\infty$ it decays as $R^{-\Delta}$. As a distribution in $t$ at large $R$ it is $\delta(t)\,R^{\Delta-1}$ plus terms that fall as $R^{-\Delta}$, so its source at the mouth in \eqref{eq:AB} is $\delta(t)$. For a source that acts during a finite time the field is
\begin{equation}\label{eq:Psisourced}
    \Psi_{\ell m}(v,R)=\int_{-\infty}^{v}dv'\,\mathcal{K}(v-v',R)\,A_{\ell m}(v')\,.
\end{equation}

The radius enters \eqref{eq:Kext} only through $Rv$, so each transverse derivative brings a factor $v$. Thus, on the horizon,
\begin{equation}\label{eq:Kder}
    \partial_R^N\mathcal{K}\big|_{R=0}=h_N\,\Theta(v)\,v^{N-\Delta}\,,\qquad h_N=\frac{(-1)^N\,\Gamma(\Delta+N)\,\sin\pi\Delta}{2^{\Delta+N-1}\sqrt{\pi}\,\Gamma\!\left(\Delta-\frac12\right)}\,,
\end{equation}
and at late times the field \eqref{eq:Psisourced} has
\begin{equation}\label{eq:Psider}
    \partial_R^N\Psi_{\ell m}\big|_{R=0}\simeq h_N\,v^{N-\Delta}\int_{-\infty}^{\infty}dv'\,A_{\ell m}(v')\,.
\end{equation}
For non-integer $\Delta$ the factor $\sin\pi\Delta$ does not vanish, and a generic source then makes all the derivatives with $N>\Delta$ grow without bound, the first one with $N=\lceil\Delta\rceil$. This is the Aretakis instability, here with no conserved quantity on the horizon.

The source enters \eqref{eq:Psider} only through its integral, which is its component at zero frequency. When this vanishes, the next term, set by the first moment of the source, grows more slowly by one power of $v$, so the growth starts one derivative later. Data supported away from the horizon of extremal Reissner--Nordstr\"om, which have a vanishing Aretakis constant, behave in the same way \cite{Aretakis:2012bm,Lucietti:2012xr}.

\paragraph{Boundary response.}
The same solution \eqref{eq:Psisourced} gives the response at the mouth. There, at fixed $t>0$, \eqref{eq:Kext} falls as $R^{-\Delta}t^{-2\Delta}$, and through the first line of \eqref{eq:GRcl} its coefficient gives the conformal response \eqref{eq:GRearly}, now at all times.

The growth on the horizon and the decay at the mouth are therefore the two ends of a single profile, $(2+y)^{-\Delta}$ with $y=Rv$. On the horizon, at $y=0$, its derivatives give the growth. At the mouth, as $y\to\infty$, its falloff $y^{-\Delta}$ gives the decay \cite{Gralla:2018xzo,Gralla:2018xoz}.

The branch point of the response at zero frequency, \eqref{eq:GRext}, does not imply the growth by itself, which also depends on how the field varies along the radius. In the extremal throat this radial profile is fixed by the scaling symmetry, together with the wave equation and the conditions at the horizon and at the mouth. So the branch point at the mouth comes with growth on the horizon.

\paragraph{Integer $\Delta$.}
When $\Delta$ is an integer the factor $\sin\pi\Delta$ vanishes, and so does $\mathcal{K}$ for $v>0$, in accord with section~\ref{subsec:classical_QNMs}, where the response at the mouth is instantaneous. A source at the mouth then leaves no field in the classical throat once it is off, and does not excite the growth.

In the isolated throat the growth then needs a non-zero Aretakis constant, \ie data on the horizon \cite{Aretakis:2012ei,Lucietti:2012xr}. The solution that carries it is still self-similar: for $\Delta=1$ the late-time field near the horizon is proportional to $[v(2+Rv)]^{-1}$, with the Aretakis constant as coefficient \cite{Hadar:2018izi}. In the complete black hole, data away from the horizon also make the transverse derivatives grow \cite{Aretakis:2012bm,Lucietti:2012xr}.

\subsection{The near-extremal throat}
\label{subsec:Aretakis_NE}

\paragraph{Transient growth.}
At finite temperature the scalings are no longer a symmetry of the black hole patch of AdS$_2$, but this patch is part of the Poincar\'e one \cite{Hadar:2018izi}. In the coordinates
\begin{equation}\label{eq:Poincare}
    \tilde v=\frac{e^{2\pi Tv}-1}{2\pi T}\,,\qquad \tilde R=e^{-2\pi Tv}R\,,
\end{equation}
the metric \eqref{eq:EF} takes its extremal form, with the same future horizon, and a source $\delta(t)$ at the mouth is the same in both coordinates. The propagator is therefore \eqref{eq:Kext} with $\tilde v$, $\tilde R$ in place of $v$, $R$. With $\partial_R=e^{-2\pi Tv}\partial_{\tilde R}$ at fixed $v$, on the horizon
\begin{equation}\label{eq:KderT}
    \partial_R^N\mathcal{K}\big|_{R=0}=h_N\,\Theta(v)\left(\frac{1-e^{-2\pi Tv}}{2\pi T}\right)^{N-\Delta}e^{-2\pi T\Delta v}\,.
\end{equation}
When $2\pi Tv\ll1$ this is \eqref{eq:Kder}. When $2\pi Tv\gg1$ every derivative decays as $(2\pi T)^{\Delta-N}e^{-2\pi T\Delta v}$. The derivatives with $N>\Delta$ grow as in the extremal throat, reach a maximum of order $(2\pi T)^{\Delta-N}$ at
\begin{equation}\label{eq:vmax}
    v=\frac{1}{2\pi T}\ln\frac{N}{\Delta}\,,
\end{equation}
and then decay \cite{Gralla:2016sxp,Zimmerman:2016qtn}. A source of finite duration gives the same transient. The redshift has turned the instability into a transient, whose duration and height diverge as $T\to0$.

\paragraph{Boundary response and quasinormal modes.}
At the mouth the propagator decays as $R^{-\Delta}$, as at extremality, but with $t^{-2\Delta}$ replaced by $(\pi T/\sinh\pi Tt)^{2\Delta}$. Its coefficient is then the thermal response \eqref{eq:GRclt}. This is conformal at early times, $t\ll1/T$, and at late times it decays as $e^{-2\pi T\Delta t}$, at the same rate as the horizon derivatives.

This rate is the damping of the fundamental quasinormal mode. To see how the other modes enter the horizon derivatives, we take the Fourier transform of \eqref{eq:KderT}. With $e^{-2\pi Tv}$ as the variable of integration it becomes an Euler beta integral, and
\begin{equation}\label{eq:KNhat}
    \int_0^\infty dv\,e^{i\omega v}\,\partial_R^N\mathcal{K}\big|_{R=0}=h_N\,\Gamma(N+1-\Delta)\,(2\pi T)^{\Delta-N-1}\,\frac{\Gamma\!\left(\Delta-\frac{i\omega}{2\pi T}\right)}{\Gamma\!\left(N+1-\frac{i\omega}{2\pi T}\right)}\,,
\end{equation}
where the integral converges, for $N>\Delta-1$ and $\mathrm{Im}\,\omega>-2\pi T\Delta$, and by meromorphic continuation elsewhere. The numerator is the Gamma function whose poles give the quasinormal frequencies \eqref{eq:zdm} of the classical response. For non-integer $\Delta$ the denominator cancels none of them, and summing their residues gives \eqref{eq:KderT} as a sum over modes,
\begin{equation}\label{eq:KNmodes}
    \partial_R^N\mathcal{K}\big|_{R=0}=h_N\,(2\pi T)^{\Delta-N}\sum_{n=0}^\infty\frac{\Gamma(\Delta-N+n)}{\Gamma(\Delta-N)\,n!}\,e^{-2\pi T(\Delta+n)v}\,,
\end{equation}
which is also the binomial expansion of \eqref{eq:KderT}. The response at the mouth is the same sum with $\Gamma(2\Delta+n)/\Gamma(2\Delta)$ in place of $\Gamma(\Delta-N+n)/\Gamma(\Delta-N)$. The mouth and the horizon thus share the classical quasinormal frequencies, with different residues.

For a source that acts during a finite time, \eqref{eq:Psisourced} gives, once the source is off,
\begin{equation}\label{eq:PsimodesT}
    \partial_R^N\Psi_{\ell m}\big|_{R=0}=h_N\,(2\pi T)^{\Delta-N}\sum_{n=0}^\infty\frac{\Gamma(\Delta-N+n)}{\Gamma(\Delta-N)\,n!}\,e^{-2\pi T(\Delta+n)v}\int_{-\infty}^{\infty}dv'\,e^{2\pi T(\Delta+n)v'}A_{\ell m}(v')\,.
\end{equation}
Each mode is weighted by its own moment of the source, and the fundamental one, $n=0$, sets the late decay. This common spectrum follows from the classical dynamics of the throat. It could not be inferred from the response at the mouth alone, without the radial structure of the solution.

\paragraph{Extremal limit.}
At fixed $v$, $T\to0$ turns \eqref{eq:KderT} into \eqref{eq:Kder}, with its unbounded growth, while at fixed $T$, $v\to\infty$ makes every derivative decay. The two limits do not commute. In frequency, the spacing $2\pi T$ of the modes goes to zero, and at fixed $\omega$ \eqref{eq:KNhat} tends to $h_N\,\Gamma(N+1-\Delta)\,(-i\omega)^{\Delta-N-1}$. This is the transform of \eqref{eq:Kder}, with a branch point at $\omega=0$, as the response tends to \eqref{eq:GRext}. The growth is a collective effect of the condensing tower: no single mode grows.

\subsection{The quantum throat}
\label{subsec:Aretakis_qu}

In the classical throat the growth rests on three features. The first is the tower of modes, whose damping vanishes and which condenses into a branch point at zero frequency. The second is the scaling symmetry, which makes the solution self-similar and ties the decay at the mouth to the growth on the horizon. The third is a classical horizon, on which the transverse derivatives are taken. The quantum throat has none of them.

\paragraph{No zero-damped modes.}
At fixed $E_b$ the tower does not condense. The poles \eqref{eq:qqnm} have real parts set by $E_b$. As $E\to0$ they approach the real axis at the fixed frequencies \eqref{eq:poleslow}, and they stay apart.

Nor does their vanishing damping produce long-lived modes. A pole is a quasinormal mode only while it satisfies \eqref{eq:QNMcond}, \ie while the state it leads to lies within the continuum, and its damping then exceeds $\Delta^2E_b$, by \eqref{eq:dampbound}. The poles that come close to the real axis have all been pruned.

Below $E=\Delta^2E_b/2$ no quasinormal modes are left, although both infinite families of poles are still there. The response is then the integral around the cuts, \eqref{eq:GRonlycut}. It passes from the conformal decay to the tail with no exponential stage. At late times it is set by the thresholds at the bottom of the spectrum, not by a ringdown that lasts longer and longer.

So no mode of the quantum throat is long-lived: all decay faster than $\Delta^2E_b$, and at late times the response is the tail \eqref{eq:tail}. In the classical throat, instead, the condensing tower comes with the growth on the horizon. Its absence does not settle the fate of the instability, which is not modal, but it removes one of its ingredients.

\paragraph{Broken scale invariance.}
The scale $E_b$ breaks the scaling symmetry of the extremal throat. The quantum response is therefore not self-similar, and the link between decay at the mouth and growth on the horizon that section~\ref{subsec:Aretakis_ext} drew from the symmetry does not carry over. Lowering the energy does not restore it. The limit $E\to0$ keeps $E_b$ fixed, and it differs from the classical extremal limit, which takes $E_b\to0$ before $T\to0$ (section~\ref{subsec:limits}).

In time, the response is conformal only while $t\ll1/E_b$, with corrections of relative order $E_\star t$, as in \eqref{eq:GRearly2}. At $t\sim1/E_b$ the thresholds take over, and the late response is the tail \eqref{eq:tail}. In the particle picture of section~\ref{sec:particle}, the conformal stage is the motion deep inside the wall, where the energy of the particle plays no role. It ends at $t\sim1/E_b$, when the wavelength of the particle grows to the scale of the wall.

The tail is also a power law, but not a conformal one. The decay $t^{-2\Delta}$ comes from the scaling symmetry, which ties it to the growth on the horizon. The $t^{-3/2}$ tail comes from the square-root edge of the density of states: its exponent does not depend on $\Delta$, and it oscillates at the threshold frequency $|E-E_{0,\ell}|$, not at that of any pole.

So a power law at the mouth no longer signals growth on the horizon, not even for $\Delta=3/4$, when both decay as $t^{-3/2}$.

\paragraph{No classical horizon.}
The classical instability concerns transverse derivatives on a sharply defined horizon in a fixed geometry. Beyond $t\sim1/E_b$ the throat fluctuates strongly and has neither. Transverse derivatives on the horizon then lose their classical meaning: they are ill-defined, or depend on a notion of horizon, or of infalling observer, dressed to the quantum fluctuations \cite{Franken:2026bff,Franken:2026lgp}. We make no such choice. Our statements concern the response at the mouth and the classical throat up to times of order $1/E_b$,
beyond which the growth law \eqref{eq:Psider} does not extend.

\paragraph{How far the growth goes.}
Up to that time the growth does take place. When $E\lesssim E_b$, the fluctuations of the Schwarzian mode are small at times $t\ll1/E_b$, and over these times the throat is the classical extremal one: its response is conformal, the field in it is \eqref{eq:Psisourced} with the propagator \eqref{eq:Kext}, and its transverse derivatives grow as in \eqref{eq:Psider}. As $v$ approaches $1/E_b$ they reach values of order
\begin{equation}\label{eq:Hmax}
    \partial_R^N\Psi_{\ell m}\big|_{R=0}\sim h_N\,E_b^{\Delta-N}\int_{-\infty}^{\infty}dv'\,A_{\ell m}(v')\,.
\end{equation}

When $E\gg E_b$ the redshift stops the derivatives earlier, at $v\sim1/T$, at smaller values, of order $T^{\Delta-N}$. As the black hole approaches extremality at fixed $E_b$, the height of the transient therefore stops growing as $T^{\Delta-N}$ and levels off at $E_b^{\Delta-N}$. For integer $\Delta$ the same holds for the self-similar solution that carries the Aretakis constant.

\medskip

All the ingredients of the classical Aretakis instability are thus lost in the quantum throat by $t\sim1/E_b$. The growth on the horizon reaches only a finite value while the throat is classical, and after that the response at the mouth decays with the $t^{-3/2}$ tail instead of $t^{-2\Delta}$. In this sense the Aretakis phenomenon does not survive in the quantum black hole. This conclusion needs no quantum definition of the horizon and, like the rest of our results, it holds in the disk approximation.

\section{Concluding remarks: quantum-consistent relaxation}
\label{sec:concl}

We began with a puzzle: classically, a perturbation near extremality can decay through an infinite tower of quasinormal modes, at rates far beyond the energy $E$ that the black hole has above extremality. It is of the same kind as the puzzle of \cite{Preskill:1991tb}, where the semiclassical analysis of Hawking emission predicted quanta that would take the black hole below extremality. In both cases there are no black hole states to support these phenomena, and both puzzles are resolved by the same physics: the quantum fluctuations of the throat geometry.

For relaxation, the resolution is the pruning of the tree.
The spectrum of the throat has a bottom, and a black hole of energy $E$ keeps only the finitely many pairs \eqref{eq:qqnm} that have
\begin{equation}
0\leq n<\sqrt{\frac{2E}{E_b}}-\Delta\,,
\end{equation}
so that their energies lie above the bottom. In the same way, an atom has a ground state, so an excited level has only finitely many levels below it to decay into. In both systems the bottom also turns the late decay into a power law, which for a black hole is different than the one from scale invariance in the extremal limit. A Schwarzschild black hole also has an infinite tower of overtones, but the bottom of its spectrum lies at the Planck scale, so only Planck-scale physics could cut the tower off. For a black hole near extremality, the tower is instead cut off at low energies, of order $E_b$, where the physics is under control.

What we find is an instance of a well-known feature of quantum mechanics: the semiclassical and long-time limits do not commute \cite{Berry:2001}. Since $E_b\propto\hbar$, the late-time behaviour of the classical extremal throat requires taking $\hbar\to0$ before $t\to\infty$, and at fixed $E_b$ it fails.

\paragraph{Mode stability.}
The quantum black hole is mode stable, as is the classical one \cite{Whiting:1988vc,Shlapentokh-Rothman:2013hza}. The reason is simple: the Hamiltonian of the throat is Hermitian, so the black hole states have real energies, and the survival amplitude obeys the bound \eqref{eq:abound}, as it would for any Hermitian Hamiltonian. The response decays as the components of the superposition dephase. The complex quasinormal frequencies describe this decay: they are frequencies of modes of the probe response, not energies of black hole states. Each quasinormal mode has the classical damping rate, but it is pruned before this rate becomes small, so that all of them decay at rates larger than $\Delta^2E_b$. Classically, instead, the damping vanishes as extremality is approached.

Moreover, below $E=\Delta^2E_b/2$ the throat has no quasinormal modes, so it cannot be excited into any resonance. Transitions still happen, but they spread smoothly over the continuum. For a massless scalar, the modes that do exist hardly ring, and the only sustained oscillation in the response is that of the late-time tail.

\paragraph{Quantum relaxation vs.\ Hawking emission.}
The quantum black hole still decays by Hawking emission, but this is a different process. Our response contains absorption and stimulated emission, and it decays as the excited state $|\psi\rangle$ dephases among black hole states. Classically, this is the perturbation falling through the horizon; only a small part leaks out through the mouth as classical radiation. Hawking emission is spontaneous: it needs the field outside to be quantized, and it takes the black hole to lower energies as the quantum escapes to infinity.\footnote{We consider only neutral fields. Black holes of large charge emit charged particles at exponentially suppressed rates \cite{Brown:2024ajk}.} It involves the same matrix elements \eqref{eq:matel} and the same thresholds, which resolve both puzzles: the black hole can neither emit quanta nor sustain modes that would take it below the bottom of the spectrum.

Hawking emission is also much slower than the relaxation. At frequencies $\omega\ll1/r_0$ the throat couples to the exterior radiation only through the same greybody factor that limits the classical leak, which brings powers of $\omega r_0$. When $E\sim E_b$ the emitted quanta have $\omega\sim E_b$, so $\omega r_0\sim1/S_0$, and for an s-wave scalar (the least suppressed channel) the black hole states acquire a width
\begin{equation}\label{eq:GammaH}
    \Gamma_H\sim E_b\,(\omega r_0)^2\sim\frac{E_b}{S_0^2}\ll E_b\,.
\end{equation}
The emission is much slower at lower energies, and slower still for photons and gravitons \cite{Brown:2024ajk}, but only by powers of $S_0$ and $E/E_b$. By the time the black hole emits a quantum, all of the relaxation is over except the tail.

The tail relies on the black hole staying in the state $|E\rangle$, whose sharp energy sets its frequency, $|E-E_{0,\ell}|$. Coupled to the radiation outside, the black hole states are no longer stationary, and they decay by emission at rates of order $\Gamma_H$. The survival amplitude, which requires the black hole to return to $|E\rangle$ with nothing emitted, is then damped at the same rate. Once the black hole has emitted, it is entangled with the quantum, so it is in a mixed state, with its energy spread over the range of energies of the quantum and in general with nonzero spin. The tails of these components oscillate at different frequencies and dephase, so together they decay faster, as for the thermal states in \cite{Emparan:2026relaxII}.

The $t^{-3/2}$ tail therefore holds for $1/E_b\ll t\ll1/\Gamma_H$. In asymptotically flat space it is Hawking emission that ends it, long before the discreteness of the spectrum would, since $\Gamma_H$ is suppressed only by powers of $S_0$.
In asymptotically AdS with reflecting boundary conditions, instead, the field outside the throat has no states far below $1/L$, so a black hole with $E\ll1/L$ cannot emit at all. Its tail lasts until the discreteness of the spectrum sets in.

\paragraph{Backreaction and extremal horizons.}
Our treatment crucially includes the backreaction of the probe on the quantum state of the throat, through the recoil that prunes the modes. However, it does not include a backreaction large enough to change the classical geometry, which already classically raises the question of what becomes of the Aretakis growth. Dynamical extremal Reissner--Nordstr\"om black holes can form \cite{Murata:2013daa,Gelles:2026bix}, whose metric approaches that of the static one while the field still grows on the horizon in the manner of Aretakis. At late times their throat is a dynamical AdS$_2\times S^2$ governed by JT gravity \cite{Porfyriadis:2025pov}, which is where the Schwarzian fluctuations of section~\ref{sec:quantum_setup} enter. We expect that $E_b$ cuts off the growth there too, as we have argued for linear perturbations, but this requires a separate analysis.

Extremal horizons suffer from a related classical pathology: in AdS, almost all extremal black holes have singular horizons \cite{Horowitz:2022mly}. These singularities also originate in the scaling of the throat. At nonzero temperature the tidal forces on the horizon are finite, but grow as a negative power of $T$, and it is natural to ask whether $E_b$ caps them, as it caps the Aretakis growth. Which effect sets in first depends on the size of the deformation: for small ones it is the Schwarzian, while for larger ones, as in Kerr--Newman black holes of realistic charge, it is higher-derivative corrections~\cite{Horowitz:2024dch}.

\paragraph{Rotating black holes.}
Near-extremal Kerr black holes also have a throat whose low-energy quantum dynamics is that of the Schwarzian mode, together with a rotational mode \cite{Kapec:2023ruw,Rakic:2023vhv,Luo:2026epp}. Our results rely on the square-root edge of the Schwarzian density of states, which gives the $t^{-3/2}$ tail, and on the matrix elements \eqref{eq:matel}, which give the pairs of poles, the recoil and the pruning. We therefore expect them to hold for probes of real conformal dimension, among them the axisymmetric ones that show the Aretakis instability of extremal Kerr \cite{Lucietti:2012sf}. The pruning condition \eqref{eq:prune} does not involve the rotational energy, so it should hold unchanged. Many non-axisymmetric probes, however, have complex conformal dimension, and they give the fastest Aretakis growth \cite{Casals:2016mel,Gralla:2016sxp}. Our analysis does not apply to them directly.

\paragraph{Open questions and outlook.}
At late times the response at the mouth decays as $t^{-3/2}$. Whether this drives some other kind of growth is a question about what an observer falling into the quantum throat would measure~\cite{Franken:2026bff,Franken:2026lgp}.

Nor is the relaxation that we have described fully quantum-consistent yet. Our response decays forever, but a black hole has a discrete spectrum, and its response must eventually recur, if it does not evaporate before then. How the tail gives way to the recurrences is a much harder question, bound up with the unitarity of black holes \cite{Maldacena:2001kr}.

Our results are for black holes of fixed energy. Thermal black holes relax differently, with a blurring of their quasinormal modes, as we show in a companion paper \cite{Emparan:2026relaxII}.

\section*{Acknowledgments}
We thank Paolo Arnaudo, Panos Betzios, Anna Biggs, Bingxin Lao, Juan Maldacena, Christiana Pantelidou, Olga Papadoulaki, Harvey S. Reall, Mikel Sanchez-Garitaonandia, Marija Toma\v{s}evi\'c, Mykhaylo Usatyuk, Herman Verlinde, and Ben Withers for conversations.
RE and ST are supported by MICIU grant PID2025-169343NB-C21, and grant CEX2024-001451-M funded by MICIU/AEI/10.13039/501100011033. The project that gave rise to these results received the support of a fellowship from ``la Caixa” Foundation (ID 100010434), awarded to ST, with code LCF/BQ/DI24/12070002.
S.M.~acknowledges the support of the STFC UK grant ST/X000753/1.

\paragraph{Use of AI tools.} After the authors had carried out the main analysis and calculations, Anthropic’s Claude (Claude Opus 5.5 and Claude Fable 5.2) was used to refine arguments, check calculations, and help with drafting and auditing. OpenAI's ChatGPT (GPT-6 Astra) was used for further independent cross-checks and audits. The exact formulation of the particle picture in appendix~\ref{app:particle} was developed mostly by Claude and then verified by the authors, who take full responsibility for the content of this paper.

\appendix
\section{Exact form of the Stieltjes transform}
\label{app:Stieltjes}

The function $\mathcal{R}$ of \eqref{eq:Rdef} gives $\mathcal{S}$ on both sheets through \eqref{eq:SIR}. Here we write it as a sum over the poles of $\mathsf{F}$.

Take first $0<\Delta<1/2$, where \eqref{eq:SDelta} converges without a regulator. With $p=k'^2$ and the oddness of $\mathsf{F}$,
\begin{equation}\label{eq:SIeven}
    \mathcal{S}^{\rm I}(a)=\int_{-\infty}^{\infty}dk'\,\frac{k'\,\mathsf{F}(k')}{k'^2-a}\,.
\end{equation}
We close the contour in the upper half-plane, along arcs that pass between the poles of $\mathsf{F}$, where $|\mathsf{F}|$ is of order $|k'|^{4\Delta-2}$ and the arcs do not contribute. The pole of the kernel at $k'=u$, with $u$ the root \eqref{eq:udef}, gives $i\pi\,\mathsf{F}(u)$. The poles of $\mathsf{F}$ in the upper half-plane are at $-u_p$, with $u_p=\pm k-i(\Delta+n)$ the poles \eqref{eq:upoles}, and $\mathsf{F}$ has the same residue at $-u_p$ as at $u_p$ because it is odd. With $u_n=k-i(\Delta+n)$, these are the residues of \eqref{eq:rn}: $u\,\mathrm{Res}\,\mathsf{F}$ is $r_n^+/(2\pi\mathcal{N})$ at $u_n$, and $-r_n^-/(2\pi\mathcal{N})$ at $-u_n^*$, because $\mathsf{F}(u^*)=\mathsf{F}(u)^*$. Then, by \eqref{eq:SIR},
\begin{equation}\label{eq:Rpoles}
    \mathcal{R}(a)=\frac{i}{\mathcal{N}}\sum_{n=0}^{\infty}\left[\frac{r_n^-}{(u_n^*)^2-a}-\frac{r_n^+}{u_n^2-a}\right].
\end{equation}
So $\mathcal{R}$ has simple poles at $a=u_n^2$ and $a=(u_n^*)^2$, and no other singularities. On the first sheet $i\pi\,\mathsf{F}\big(u(a)\big)$ has poles at the same points and cancels them; on the second sheet the two add.

The terms of \eqref{eq:Rpoles} are of order $n^{4\Delta-3}$, so the sum converges for $\Delta<1/2$, like the integral. For larger $\Delta$ the $J$-th derivatives of both with respect to $a$, with $J=\lfloor2\Delta\rfloor$, converge, and the argument holds for them. Then \eqref{eq:Rpoles} holds up to a polynomial in $a$ of degree $J-1$, and it converges once each term has its first $J$ Taylor terms at $a=0$ subtracted,
\begin{equation}\label{eq:Rsub}
    \frac{1}{u_n^2-a}\;\longrightarrow\;\frac{1}{u_n^2-a}\left(\frac{a}{u_n^2}\right)^{J},
\end{equation}
and likewise for $(u_n^*)^2$. By \eqref{eq:apm} the polynomial adds to \eqref{eq:GRStieltjes} an even polynomial in $\omega$, which is a contact term. This holds for integer $\Delta$ as well.

For $\Delta<1/2$ the sum over $n$ is a hypergeometric function of unit argument,
\begin{equation}\label{eq:R3F2}
    \mathcal{R}(a)=-\pi^2\,\frac{\sinh2\pi(k-i\Delta)}{\sinh2\pi k}\,\frac{\Gamma(2\Delta)\,\Gamma(2\Delta+2ik)}{\Gamma(1+2ik)}\sum_{\pm}\frac{1}{z_\pm}\;{}_3F_2\!\left(\begin{matrix}2\Delta,\;2\Delta+2ik,\;z_\pm\\ 1+2ik,\;1+z_\pm\end{matrix};1\right)+(k\to-k)\,,
\end{equation}
with $z_\pm=\Delta+ik\pm i\sqrt{a}$, for either root of $a$.

\section{The particle picture of the relaxation from Liouville quantum mechanics}
\label{app:particle}

Here we derive the picture of section~\ref{sec:particle} from the exact solution of the Liouville quantum mechanics \eqref{eq:Liouville} \cite{Bagrets:2016cdf,Harlow:2018tqv,Kitaev:2018wpr}.

\paragraph{Eigenstates.}
With the coefficient of the potential in \eqref{eq:Liouville}, which fixes the additive constant in $\lambda$, the eigenstates of energy \eqref{eq:Ek} are
\begin{equation}\label{eq:psik}
\chi_k(\lambda)=K_{2ik}\big(4e^{-\lambda/2}\big)\,,\qquad k\geq0\,,
\end{equation}
where $K$ is the modified Bessel function. Inside the wall they decay as $\exp(-4e^{-\lambda/2})$. Far from it they are standing waves,
\begin{equation}\label{eq:psiasym}
\chi_k(\lambda)\simeq\frac12\Big[\Gamma(2ik)\,2^{-2ik}\,e^{ik\lambda}+\Gamma(-2ik)\,2^{2ik}\,e^{-ik\lambda}\Big]\,,
\end{equation}
of amplitude $|\Gamma(2ik)|$, with
\begin{equation}\label{eq:Gamma2ik}
|\Gamma(2ik)|^{-2}=\frac{2k}{\pi}\,\sinh2\pi k\,.
\end{equation}
They are orthogonal, with the Schwarzian density of states in the measure,
\begin{equation}\label{eq:orth}
\int_{-\infty}^{\infty}d\lambda\,\chi_k(\lambda)\,\chi_{k'}(\lambda)=\frac{\pi^2}{4k\sinh2\pi k}\,\delta(k-k')\,.
\end{equation}
The wave $e^{-ik\lambda}$ moves towards the wall and $e^{ik\lambda}$ away from it, so the reflection coefficient is \cite{Harlow:2018tqv}
\begin{equation}\label{eq:reflection}
\sR(k)=2^{-4ik}\,\frac{\Gamma(2ik)}{\Gamma(-2ik)}\,,
\end{equation}
which tends to $-1$ when $k\to0$: the wall reflects a slow particle as a hard wall would.

\paragraph{The response.}
The Euclidean two-point function of $O_\Delta$ is a path integral on the disk with the two insertions on its boundary. Cutting the disk along the geodesic that joins them leaves on each side a wave function of its length, with energy $E'$ on the boundary segment between the insertions and $E$ on the rest, and the probe propagates along the geodesic with amplitude $e^{-\Delta\lambda}$ \cite{Yang:2018gdb,Mertens:2022irh}. The squared matrix element of $O_\Delta$ in the weight \eqref{eq:weight} is therefore a matrix element, not squared, of $e^{-\Delta\lambda}$,
\begin{equation}\label{eq:matel2}
\int_{-\infty}^{\infty}d\lambda\,\chi_k(\lambda)\,e^{-\Delta\lambda}\,\chi_{k'}(\lambda)=\frac{\gamma_\Delta(k,k')}{2^{2\Delta+2}\,\Gamma(2\Delta)}\,,
\end{equation}
with $\gamma_\Delta$ the product \eqref{eq:gammaDelta}. Comparing with \eqref{eq:matel},
\begin{equation}\label{eq:matelL}
\sum_{m=-\ell}^{\ell}\big|\langle E',\ell,m|O_{\Delta\ell m}|E\rangle\big|^2=8\,e^{-S_0}E_b^{2\Delta}\int_{-\infty}^{\infty}d\lambda\,\chi_k(\lambda)\,e^{-\Delta\lambda}\,\chi_{k'}(\lambda)\,.
\end{equation}
In states of spin $\ell$ the rotation adds the energy $E_{0,\ell}$, so the excited states have energies
\begin{equation}\label{eq:Eprime}
E'=E_{0,\ell}+\frac{E_b}{2}\,k'^2\,,
\end{equation}
and with the density \eqref{eq:rhoj} the survival amplitude \eqref{eq:survival} is
\begin{equation}\label{eq:Aliouville}
\mathsf{G}(t)=(2\ell+1)\,E_b^{2\Delta}\,e^{-i(E_{0,\ell}-E)t}\int_0^\infty dk'\,\frac{4k'\sinh2\pi k'}{\pi^2}\,e^{-iE_bk'^2t/2}\int_{-\infty}^{\infty}d\lambda\,\chi_k(\lambda)\,e^{-\Delta\lambda}\,\chi_{k'}(\lambda)\,.
\end{equation}
The measure in $k'$ is that of the completeness relation that follows from \eqref{eq:orth}, so the integral over $k'$ is a propagator of the particle. Deep in the wall the eigenstates hardly depend on their energy. Inserting $\chi_{k'}(\lambda_0)/\chi_k(\lambda_0)$ in \eqref{eq:Aliouville}, with $\lambda_0\to-\infty$,
\begin{equation}\label{eq:Apropagator}
\mathsf{G}(t)=(2\ell+1)\,E_b^{2\Delta}\,e^{-i(E_{0,\ell}-E)t}\lim_{\lambda_0\to-\infty}\int_{-\infty}^{\infty}d\lambda\;\chi_k(\lambda)\,e^{-\Delta\lambda}\;\frac{\langle\lambda|e^{-iH_{\rm Sch}t}|\lambda_0\rangle}{\chi_k(\lambda_0)}\,.
\end{equation}
This is the picture of section~\ref{sec:particle}: the particle starts at $\lambda_0$, the geodesic of zero length with which the boundary segment begins, evolves for a time $t$, and overlaps with the black hole state through the factor $e^{-\Delta\lambda}$. $H_{\rm Sch}$ is real, so the propagator is symmetric in its two arguments, and \eqref{eq:Apropagator} is also the amplitude at $\lambda_0$ of the state $e^{-\Delta\lambda}\chi_k$ evolved forward for the same time $t$: the amplitude can be read from either end of the geodesic.

At finite $\lambda_0$ the eigenstates retain a dependence on their energy. For $z_0=4e^{-\lambda_0/2}\gg1$, the uniform asymptotics of the Bessel function give
\begin{equation}
\frac{\chi_{k'}(\lambda_0)}{\chi_k(\lambda_0)}=\exp\!\Big[-\tfrac12\big(k'^2-k^2\big)\,e^{\lambda_0/2}\Big]\left[1+\mathcal{O}\!\left(\frac{k'^2}{z_0^2}\right)+\mathcal{O}\!\left(\frac{k'^4}{z_0^3}\right)\right]
\end{equation}
for $k'^2$ up to order $z_0$, beyond which both sides are exponentially small. This is the Euclidean regulator $e^{-\epsilon(E'-E)}$ of \eqref{eq:areg}, with
\begin{equation}\label{eq:epslambda0}
\epsilon=\frac{e^{\lambda_0/2}}{E_b}\,,
\end{equation}
up to a factor $e^{\epsilon E_{0,\ell}}$ that tends to $1$.

\paragraph{Classical trajectories.}
In the classical limit, $E_b\to0$ at fixed temperature, $k=2\pi T/E_b$ is large. For $k\gg\Delta^2$, \eqref{eq:Apropagator} is dominated by a classical trajectory, which must have the energy $E$ of the state $\chi_k$ and start at $\lambda_0\to-\infty$. The rotation plays no role in it: $E_{0,\ell}$ enters \eqref{eq:Aliouville} only through the factor $e^{-iE_{0,\ell}t}$. No real trajectory starts at $\lambda_0$: a particle of energy $E$ is stopped by the wall. The real trajectory of energy $E$ is the bounce
\begin{equation}\label{eq:bounce}
e^{\lambda/2}=\frac{2}{k}\cosh\frac{E_bk\,t}{2}\,,\qquad\dot\lambda=E_bk\tanh\frac{E_bk\,t}{2}\,.
\end{equation}
It describes the geodesic between the two boundaries of the two-sided black hole as the time of one of them advances. It is shortest at $t=0$, when the particle is stopped by the wall, and on it $e^{-\Delta\lambda}$ is the correlator between the two sides. The geodesic of the response has both ends on the same boundary. It follows from the bounce by a shift of half a thermal period in imaginary time, $t\to t+i/2T$, which takes one boundary to the other, and its length is complex,
\begin{equation}\label{eq:onesided}
\lambda=\bar\lambda+i\pi\,,\qquad e^{\bar\lambda/2}=\frac{2}{k}\sinh\frac{E_bk\,t}{2}\,.
\end{equation}
It starts at zero length, as required. Of the two trajectories of energy $E$ that do so, $\lambda=\bar\lambda\pm i\pi$, the regulator \eqref{eq:areg} selects the upper sign. The imaginary part gives the phase of \eqref{eq:Onorm}, and
\begin{equation}\label{eq:onesidedcorr}
E_b^{2\Delta}\,e^{-\Delta\lambda}=e^{-i\pi\Delta}\left(\frac{\pi T}{\sinh\pi Tt}\right)^{2\Delta},
\end{equation}
so \eqref{eq:Apropagator} gives $(2\ell+1)\,e^{-iE_{0,\ell}t}$ times this. For $E_{0,\ell}t\ll1$ this is the classical survival amplitude, whose imaginary part gives \eqref{eq:GRclt}. The real part $\bar\lambda$ moves with the sign of the potential reversed,
\begin{equation}\label{eq:invpot}
\frac{E_b}{2}\left(P^2-4e^{-\bar\lambda}\right)=E\,,
\end{equation}
so for this motion the wall attracts the particle, and its momentum, $|P|=k\coth(E_bkt/2)$, falls from infinity to $k$. At early times, $t\ll1/E_bk$, and at all times at extremality,
\begin{equation}\label{eq:earlyrho}
\bar\lambda\simeq2\log E_bt\,,\qquad|P|\simeq\frac{2}{E_bt}\,,\qquad E_b^{2\Delta}\,e^{-\Delta\lambda}\simeq(it)^{-2\Delta}\,:
\end{equation}
deep inside the wall the kinetic and potential energies nearly cancel, the energy $E$ plays no role, and the decay is the conformal one. The particle leaves the wall, $e^{\bar\lambda/2}=2/k$, when $\sinh(E_bkt/2)=1$, i.e., at $t\sim1/T$. At late times, expanding in powers of $e^{-E_bkt}$,
\begin{equation}\label{eq:overtonecoef}
e^{-\Delta\bar\lambda}=k^{2\Delta}\,e^{-\Delta E_bk\,t}\sum_{n\geq0}\frac{\Gamma(2\Delta+n)}{\Gamma(2\Delta)\,n!}\,e^{-nE_bk\,t}\,.
\end{equation}
This is \eqref{eq:overtones}, with its coefficients. Each power is a correction from the wall to the uniform motion, and with $E_bk=2\pi T$ these are the overtones \eqref{eq:zdm}. The bounce has the same asymptotic velocity and the same powers, but its coefficients alternate in sign and it has no conformal stage.

The trajectory describes the particle while its wavelength is short compared with the scale of the wall, $|P|\gg1$. When $k\gg1$ this holds at all times, since $|P|\geq k$. When $k\lesssim1$, $|P|\simeq2/(E_bt)$ falls to order one at $t\sim1/E_b$, before the particle leaves the wall, and quantum effects end the conformal decay before the motion becomes uniform (section~\ref{subsec:stages}). For the trajectory to give the survival amplitude, the probe must also leave the motion unchanged: its recoil energy, of order $\Delta^2E_b$, must be small compared with $T$, i.e., $k\gg\Delta^2$. The modulus of \eqref{eq:Apropagator} then agrees with that of \eqref{eq:onesidedcorr}, up to relative corrections of order $1/k^2$ at fixed $\Delta$, until the tail takes over at $t_*$ \eqref{eq:tcross}. Its phase drifts away from that of the trajectory by $\Delta^2E_bt/2-\delta_0$, with $\delta_0$ of \eqref{eq:deltan}: the recoil of section~\ref{subsec:particle_recoil}. With the factor $e^{-iE_{0,\ell}t}$, the drift with respect to the classical survival amplitude is $(\Delta^2E_b/2-E_{0,\ell})\,t-\delta_0$, so the classical result holds while $|\Delta^2E_b/2-E_{0,\ell}|\,t\ll1$, as in \eqref{eq:clwindow}. After that the phase is that of the pairs \eqref{eq:paircl}. For integer $\Delta$, where \eqref{eq:onesidedcorr} is real, this phase carries the whole response \eqref{eq:pairint}.

\paragraph{Quantum poles.}
The poles of the response come from large $\lambda$, where the integral in \eqref{eq:matel2} can diverge. The eigenfunctions are given by the convergent series
\begin{equation}\label{eq:psiseries}
\chi_k(\lambda)=\sum_{n\geq0}e^{-n\lambda}\left(b_n\,e^{ik\lambda}+b_n^*\,e^{-ik\lambda}\right),
\end{equation}
whose terms with $n\geq1$ are corrections from the wall. In \eqref{eq:matel2}, the probe $e^{-\Delta\lambda}$, the ingoing wave $e^{-ik\lambda}$ of the initial state, and the conjugate $e^{ik'\lambda}$ of the ingoing wave of the final state combine into $e^{-\Delta\lambda}e^{-ik\lambda}e^{ik'\lambda}$. This product ceases to decay at large $\lambda$ when $k'=k-i\Delta$. With the wall terms, the poles are at
\begin{equation}\label{eq:kprimepoles}
k'=k+\delta k=k-i(\Delta+n)\,,
\end{equation}
with $\delta k$ of \eqref{eq:q}. These are the poles of $\Gamma(\Delta+ik-ik')$ in \eqref{eq:gammaDelta}. The other combinations of waves give the poles of the other three Gamma functions, at $k'=-k-i(\Delta+n)$ and at the complex conjugates of both. Of these, only \eqref{eq:kprimepoles} can lie in the sector \eqref{eq:sector} that the absorption term reaches from the real axis, and it does when $k>\Delta+n$ (section~\ref{subsec:poles}). Its frequency is
\begin{equation}\label{eq:recoil}
\omega=E'-E=E_{0,\ell}-i(\Delta+n)E_bk-\frac{(\Delta+n)^2}{2}\,E_b\,,
\end{equation}
the rotational energy plus the change in energy \eqref{eq:dE}, i.e., $\omega_n^+$ in \eqref{eq:qqnm}, with the damping exactly $\Delta+n$ times the velocity $E_bk$. The partners $\omega_n^-$ come from stimulated emission: in the emission term, the pole at $u=-k-i(\Delta+n)$, in the notation of section~\ref{subsec:poles}, lies in the sector \eqref{eq:sectorem} under the same condition. The energy of the excited state, $E+\mathrm{Re}\,\omega_n^+=E_{0,\ell}+E-(\Delta+n)^2E_b/2$, lies within the continuum of spin-$\ell$ states only if $E$ exceeds the recoil, which is \eqref{eq:prune}.

\paragraph{Late-time tail.}
At late times \eqref{eq:Aliouville} is dominated by excited states with $k'\to0$. There the measure vanishes as $k'^2$ and \eqref{eq:matel2} is finite, so $\mathsf{W}\propto(E'-E_{0,\ell})^{1/2}$, as in \eqref{eq:wedge}, and the decay is the $t^{-3/2}$ of \eqref{eq:tail}. The two powers of $k'$ have separate origins. In terms of the states of unit amplitude far from the wall,
\begin{equation}\label{eq:unitamp}
\widetilde\chi_k=\frac{\chi_k}{|\Gamma(2ik)|}\,,
\end{equation}
the measure in \eqref{eq:Aliouville} is $(2/\pi)\,|\Gamma(2ik')|^{-2}dk'$, by \eqref{eq:Gamma2ik}. Expanding the propagator of \eqref{eq:Apropagator} in eigenstates, its integrand in $k'$ is, up to factors independent of $k'$,
\begin{equation}\label{eq:factor}
dk'\;\widetilde\chi_{k'}(\lambda_0)\int_{-\infty}^{\infty}d\lambda\,\chi_k(\lambda)\,e^{-\Delta\lambda}\,\widetilde\chi_{k'}(\lambda)\,,\qquad\lambda_0\to-\infty\,.
\end{equation}
A slow state of unit amplitude, reflected by the wall as by a hard wall, is of order $k'$ at any finite $\lambda$, because $|\Gamma(2ik')|^{-1}\simeq2k'$. So each factor is of order $k'$: one from the start of the geodesic, at zero length, and one from its end, where the probe acts. This is the free particle on a half-line of section~\ref{subsec:particle_tail}. When $E\gg E_b$ the weight of these states carries the factor $|\Gamma(\Delta+ik)|^4$ of \eqref{eq:Fsmall}, exponentially small by \eqref{eq:tailamp}, and the tail takes over from the exponentials only at the time $t_*$ of \eqref{eq:tcross}.

\bibliography{refs_papers}

\end{document}